\documentclass[sn-basic]{sn-jnl}

\usepackage{graphicx}%
\usepackage{multirow}%
\usepackage{amsmath,amssymb,amsfonts}%
\usepackage{amsthm}%
\usepackage{mathrsfs}%
\usepackage[title]{appendix}%
\usepackage{textcomp}%
\usepackage{manyfoot}%
\usepackage{booktabs}%
\usepackage{algorithm}%
\usepackage{algorithmicx}%
\usepackage{algpseudocode}%
\usepackage{listings}%

\usepackage{comment}

\usepackage{tabulary}

\usepackage{multicol}
\usepackage{color, colortbl}

\usepackage[table]{xcolor}
\definecolor{lightgray}{rgb}{0.90, 0.90, 0.90}

\usepackage[most]{tcolorbox}

\usepackage{wrapfig}

\usepackage{enumitem}

\renewcommand{\texttt}[1]{%
  \begingroup
  \ttfamily
  \begingroup\lccode`~=`\_\lowercase{\endgroup\def~}{\_\discretionary{}{}{}}%
  \begingroup\lccode`~=`(\lowercase{\endgroup\def~}{(\discretionary{}{}{}}%
  \begingroup\lccode`~=`)\lowercase{\endgroup\def~}{)\discretionary{}{}{}}%
  \begingroup\lccode`~=`.\lowercase{\endgroup\def~}{.\discretionary{}{}{}}%
  \begingroup\lccode`~=`|\lowercase{\endgroup\def~}{|\discretionary{}{}{}}%
  \catcode`\_=\active
  \catcode`(=\active
  \catcode`)=\active
  \catcode`.=\active
  \catcode`|=\active
  \scantokens{#1\noexpand}%
  \endgroup
}

\usepackage{dirtytalk}

\usepackage{array}

\usepackage{soul}
\soulregister\cite7
\soulregister\citep7
\soulregister\ref7
\soulregister\pageref7
\soulregister\emph7
\soulregister\texttt7

\usepackage{hyperref}								
\hypersetup{colorlinks, citecolor=black,
   		 	filecolor=black, linkcolor=black,
    		urlcolor=black}

\usepackage{subfig}

\usepackage{cleveref}

\newcommand{\rrgithub}{\url{https://github.com/BWTSE/RobotResearcher/tree/persist1.2}}
\newcommand{\rrdemo}{\url{https://bwtse.github.io/RobotResearcher}}
\newcommand{\rrdoi}{\url{https://doi.org/10.5281/zenodo.7012040}}

\newcommand{\scenariogithub}{\url{https://github.com/BWTSE/Scenarios/tree/persist1.3}}
\newcommand{\scenariodoi}{\url{https://doi.org/10.5281/zenodo.7011996}}

\newcommand{\rpgithub}{\url{https://github.com/BWTSE/Analysis/tree/v1.6}}
\newcommand{\rppresent}{\url{https://bwtse.github.io/Analysis}}
\newcommand{\rpdoi}{\url{https://doi.org/10.5281/zenodo.10252819}}

\newcommand{\datagithub}{\url{https://github.com/BWTSE/Data/tree/v1.4}}
\newcommand{\datadoi}{\url{https://doi.org/10.5281/zenodo.7011992}}

\newcommand{\githubgroup}{\url{https://github.com/BWTSE}}

\begin{document}

\title[The Broken Windows Theory Applies to
Technical Debt]{The Broken Windows Theory Applies to
Technical Debt}


\author*{\fnm{William} \sur{Levén}}\email{william@leven.id}

\author{\fnm{Hampus} \sur{Broman}}\email{namorben@gmail.com}

\author[1]{\fnm{Terese} \sur{Besker}}\email{terese.besker@ri.se}
\author[2]{\fnm{Richard} \sur{Torkar}}\email{richard.torkar@gu.se}

\affil*[1]{\orgname{RISE Research Institutes of Sweden AB}, \orgaddress{ \city{Gothenburg}, \postcode{41296}, \country{Sweden}}}

\affil[2]{\orgdiv{Computer Science and Engineering}, \orgname{Chalmers and University of Gothenburg}, \orgaddress{\city{Gothenburg}, \postcode{41296}, \country{Sweden}}}

\abstract{
    \textbf{Context}: The term \emph{technical debt} (TD) describes the aggregation of sub-optimal solutions that serve to impede the evolution and maintenance of a system. Some claim that the \emph{broken windows theory} (BWT), a concept borrowed from criminology, also applies to software development projects. The theory states that the presence of indications of previous crime (such as a broken window) will increase the likelihood of further criminal activity; TD could be considered the \emph{broken windows} of software systems.
    
    \textbf{Objective}: 
    To empirically investigate the causal relationship between the TD density of a system and the propensity of developers to introduce new TD during the extension of that system. %
    
    \textbf{Method:} The study used a mixed-methods research strategy consisting of a controlled experiment with an accompanying survey and follow-up interviews. The experiment had a total of 29 developers of varying experience levels completing system extension tasks in already existing systems with high or low TD density.
    
    \textbf{Results}: The analysis revealed significant effects of TD level on the subjects' tendency to re-implement (rather than reuse) functionality, choose non-descriptive variable names, and introduce other \emph{code smells} identified by the software tool \textsf{SonarQube}, all with at least $95\%$ credible intervals.
    
    \textbf{Conclusions}: Three separate significant results along with a validating qualitative result combine to form substantial evidence of the BWT's existence in software engineering contexts. This study finds that existing TD can have a major impact on developers propensity to introduce new TD of various types during development.
}


\maketitle

\section{Introduction}\label{sec:introduction}
Contemporary software systems often have complex and constantly evolving code bases that require continuous maintenance. \cite{Martini2018TechnicalPractice} found that developers waste as much as twenty-five percent of development time refactoring or otherwise managing previous implemented sub-optimal solutions. The culprit is \emph{technical~debt}~(TD), an accumulation of \say{not-quite-right} implementations that now serve to impede further development and maintenance. The costs incurred by TD often claim a sizeable portion of the budget of software development projects and are referred to as \emph{interest}~\citep{Ampatzoglou2015TheReview}.

What is worse, unless actively managed, the costs of TD may increase nearly exponentially with TD generating additional TD as well as interest~\citep{Martini2015TheCircles,Martini2017OnPhenomenon}. 
There are multiple possible causes of this dynamic (and they are not mutually exclusive). \cite{Hunt1999TheMaster} suggested that the most important factor is the \say{psychology, or culture, at work on a project.} They explained this by analogously applying the \emph{broken~windows~theory}~(BWT) from criminology. The BWT states that an indication of previous crimes, or even just general disorder, increases the likelihood of further crimes being committed. In their view, this holds in software development projects, where acceptance of minor defects can result in a snowball effect that causes further deterioration of not only the system, but also the culture surrounding the project. 

It is a compelling argument that a developer might think that if someone else got away with being careless, perhaps they could too. From a psychological perspective, this could be explained through the lens of \emph{norms}, where the apparent prevalence of a behavior forms a \emph{descriptive norm}, signaling that the behavior is acceptable. It could also be that it happens through unconscious mimicry of previous work, or that it is simply a result of being less enthusiastic about the system due to the TD\@.
Though this \emph{software engineering} BWT gained some traction, there appears to be no empirical evidence for it. \cite{Besker2020TheMorale} found that the existence of TD impairs the morale of software developers, lending some credence to the theory, but did not examine the tangible effects of that decline. 

This study aims to evaluate the causal relationship between the TD density of a system and the propensity of developers to introduce TD during the extension of the system. While there is some research indicating the developers often find themselves \emph{forced} to introduce TD as a result of previous suboptimal solutions~\citep{Besker2019SoftwareWork}, our interest is in situations where no significant hurdles are preventing the developer from implementing a low TD solution. In other words, our goal is to understand if they will adopt the negligent attitude towards the software that they may infer is acceptable, given its state.

The study utilizes a mixed-methods design combining a controlled experiment and a survey with semi-structured interviews. The experiment involved software practitioners completing tasks that required the extension of small existing systems, designed specifically for the experiment, half of which we had deliberately injected with TD\@. Volunteering participants were subsequently interviewed, and asked about their reason for choosing their particular solution, their general thoughts around TD management, and their experiences on the BWT in software engineering. The quantitative data from the experiment and survey were interpreted using \emph{Bayesian data analysis} and the qualitative data produced by the interviews through \emph{thematic analysis}.

Efficient management of TD requires an understanding of the dynamics that govern its growth and spread. The results of this study could be a first step toward formulating improved strategies to keep TD in check. Additionally, the experiment could be seen as a cross-disciplinary effort that could speak to the generalizability of the original Broken Windows Theory.

\subsection{Research Questions}
The goal of this study is to isolate and measure the causal effects of existing technical debt on a developer's propensity to introduce new technical debt. We condense this into the following research questions, and by distinguishing between \emph{similar} and \emph{dissimilar} TD, we ask whether developers mostly mimic TD they come in contact with or if they also introduce TD of a different character than that which was previously in the system.

\begin{enumerate}[label*=\emph{RQX.X*}, leftmargin=*]
    \item [RQ1:] How does existing technical debt affect a developer's propensity to introduce technical debt during further development of a system?
    \item [RQ2:] How does existing technical debt affect a developer's propensity to introduce \emph{similar} technical debt during further development of a system?
    \item [RQ3:] How does existing technical debt affect a developer's propensity to introduce \emph{dissimilar} technical debt during further development of a system?
\end{enumerate}

\section{The Broken Windows Theory}

The story of the BWT starts with an experiment conducted by Philip Zimbardo in 1969. He placed identical cars, with no license plates and the hoods up, in Bronx, New York, and Palo Alto, California (on the Stanford campus).

The car in New York was destroyed and stripped of parts after just three days, while the car left on the Stanford campus went untouched for over a week~\citep{Zimbardo1969}. 
Zimbardo was not investigating the BWT, rather, he wanted to observe what kind of people committed acts of vandalism and under what circumstances.
However, he did (successfully) employ the BWT when he initiated the vandalism by taking a sledgehammer to the car on the Stanford campus, after which others continued the vandalism within hours~\citep{Zimbardo1969}.

A sample size of one is hardly convincing. However, Zimbardo's experiment inspired Wilson and Kelling to write an opinion piece in the monthly magazine \say{The Atlantic}. The text, \emph{Broken Windows: The police and neighborhood safety}~\citep{WILSONGEORGEL.KELLING1982BrokenSafety}, takes the concept beyond the narrow domain of vandalism and applies it to a broader context of crime and antisocial behavior. In doing so, they created what subsequently became known as the BWT\@. 

Wilson and Kelling suggest (although they provide no evidence) that:

\begin{quote}
Window-breaking does not necessarily occur on a large scale because some areas are inhabited by determined window-breakers whereas others are populated by window-lovers; rather, one unrepaired broken window is a signal that no one cares, and so breaking more windows costs nothing. (It has always been fun.)~\citep{WILSONGEORGEL.KELLING1982BrokenSafety}
\end{quote}

\noindent
Perhaps the most prolific example of policy shaped by the BWT is that of New York City under former mayor Rudy Giuliani and police commissioner William Bratton. The strategy adopted revolved around strictly enforcing minor crimes and was followed by a 50\% drop in property and violent crime~\citep{Corman2002CarrotsWindows}. While the political leadership attributed the improvement to their broken windows policing policies, the causality of this relationship has been questioned~\citep{Thacher2004OrderReasoning, Sampson1999}.

Several more recent studies have provided empirical evidence supporting the BWT\@. \cite{Keizer2008} secretly observed a number of public areas in Groningen, to some of which they had deliberately introduced disorder, e.g., graffiti, and found that it had a significant effect on further minor crime (such as graffiti or littering), but also on more serious crime such as robberies. Additionally, a 2015 meta-study found that broken window policing strategies moderately decrease crime levels~\citep{Braga2015CanMeta-analysis}.

The BWT has also been examined from a more psychological perspective, in the context of \emph{normativity}. Several studies within the field have evaluated the effects of a littered environment on the propensity of subjects to litter themselves~\citep{Krauss1978StudiesLittering,Cialdini1990APlaces}. \cite{Cialdini1990APlaces} argue that a littered environment conveys the descriptive norm that littering is acceptable.

The first reference to the BWT in a software engineering context appears in \emph{The Pragmatic Programmer: From Journeyman to Master}, where authors Hunt and Thomas assert that the BWT holds in the software development context~\citep{Hunt1999TheMaster}. They describe the psychology and culture of a project team as the (likely) most crucial driver of \emph{software entropy} (sometimes referred to as \emph{software rot}), and the \say{broken windows,} they claim, are significant factors in shaping them~\citep{Hunt1999TheMaster}:

\begin{quote}
    One broken window---a badly designed piece of code, a poor management decision that the team must live with for the duration of the project---is all it takes to start the decline. If you find yourself working on a project with quite a few broken windows, it's all too easy to slip into the mindset of \say{All the rest of this code is crap, I'll just follow suit.}~\citep[p.~8]{Hunt1999TheMaster}
\end{quote}
\noindent
Hunt and Thomas provide no evidence to support this claim, other than referring to the 1969 Zimbardo experiment~\citep{Zimbardo1969} and an (unsupported) assertion that anti-broken window measures have worked in New York City~\citep{Hunt1999TheMaster}. The only software development-specific arguments come in the form of anecdotal evidence~\citep{Hunt1999TheMaster}. 

Since its inception, the BWT in software engineering has mostly been discussed on online blogs and, sometimes using different terminology, in popular books such as \emph{Clean Code}~\citep{Martin2014CleanCraftmanship} with the occasional mention in a scientific paper~\citep{Sharma2015ChallengesPerspective}. Section~\ref{sec:related_literature} further explores the prevalence of the BWT in software engineering research.

\section{Technical Debt}

\begin{quote}
Shipping first time code is like going into debt. A little debt speeds development so long as it is paid back promptly with a rewrite\ldots The danger occurs when the debt is not repaid. Every minute spent on not-quite-right code counts as interest on that debt.
    \begin{flushright}
    \citep[p.~30]{Cunningham1992TheSystem}    
    \end{flushright} 
\end{quote} 
\noindent
Ward Cunningham (co-author of \emph{The Agile Manifesto}~\citep{ManifestoDevelopment}) coined the metaphor in his 1992 OOPSLA experience report~\citep{Cunningham1992TheSystem}. Since then, the concept of TD has gained widespread popularity and is commonly applied, not just to code but also to related concepts, resulting in terms such as \emph{documentation technical debt} and \emph{build technical debt}~\citep{Brown2010ManagingSystems}.

Technical debt is now a concept used and studied academically. While there certainly are multiple definitions to choose from, in this paper, we will use the following one arrived at during the 2016 \say{Dagstuhl Seminar 16162: Managing Technical Debt in Software Engineering}:

\begin{quote}
In software-intensive systems, technical debt is a collection of design or implementation constructs that are expedient in the short term, but set up a technical context that can make future changes more costly or impossible. Technical debt presents an actual or contingent liability whose impact is limited to internal system qualities, primarily maintainability and evolvability.~\citep[p.~112]{Avgeriou2016}
\end{quote}
\noindent
Studying the propagation of TD requires a method for identifying it. There are multiple such methods, ranging from those that require more manual labor to those that rely on automated processes~\citep{Martini2015TheCircles, OlssonMeasuringThe, Besker2020TheMorale}. Some TD could be painfully apparent to developers; after all, it might impact their professional life daily. Simply asking project members could be a way of identifying some of the most egregious instances of TD\@.

Other types of TD can be a little more subtle. Manually scanning the code for these would generally be prohibitively time-consuming, hence a need for automation. Static code analysis tools see frequent use in industry and research contexts; they rely on rule sets to identify TD items, often referred to as \emph{code smells}~\citep{Alves2016}. Some such tools even attempt to estimate the time required to fix each item~\citep{Alves2016}. This category ranges from simple linting rules that can do little beyond correcting indentation to more encompassing solutions that detect security issues and deviations from language convention. 

We propose to interpret TD items as the broken windows of \emph{software engineering} as introduced by Hunt and Thomas~\cite{Hunt1999TheMaster}. In our view, several recent findings support this conceptualization. 

\cite{Besker2020TheMorale} found TD to negatively impact developer morale, establishing a link between TD and the psychological state of those working in its presence. This lowered morale is reminiscent of the dynamic described by Wilson and Kelling in their original \emph{Broken Windows} article, where they assert that general disorder will trigger the local inhabitants to adopt a general stance of \say{not getting involved}~\citep{WILSONGEORGEL.KELLING1982BrokenSafety}. This link is further supported by a study by \cite{OlssonMeasuringThe} that used the SAM (self-assessment manikin) system to map changes in affective states triggered by interaction with TD\@. A natural continuation of that research would be to examine the effects of this psychological impact, where propagation or reproduction of TD may be one of the most exciting lines of inquiry. Moreover, \cite{FERNANDEZSANCHEZ201722} have also described that besides causing low morale in development teams, TD may also cause negative effects such as developer turnover.

\section{Related Literature}
\label{sec:related_literature}
\cite{Tufano2015WhenBad} found that most code smells were introduced during the creation of an artifact or feature and that when code smells were introduced during the evolution of the project, it was usually in close proximity to preexisting code-smells. They also observed that a significant portion was introduced during refactoring activities and that developers with high workloads or tight schedules were more likely to introduce them. On the other hand, \cite{Digkas2020OnDebt} examined the change in \emph{code technical debt} (CTD) over time in a set of open-source projects and attempted to correlate the amount of CTD introduced with the workload of the development teams. They did, however, not find any correlation between the workload of the teams and the amount of new CTD introduced. Furthermore, the same study found that the introduction of CTD was mainly stable over time, with a few spikes.
In contrast, \cite{Lenarduzzi2020DoesDebt} followed the migration of a monolithic system to a micro-services system and found that for each individual micro-service, the amount of TD was very high in the beginning but decreased as the artifact matured. The study also reported that the migration reduced the rate at which TD grew in the system but that, over time, TD still grew.

\cite{Chatzigeorgiou2014InvestigatingSystems} as well as \cite{Digkas2020CanDensity} found that TD is mainly introduced when adding new features to a code base and that as they are not removed, they accumulate over time. Similarly, \cite{Bavota2016ADebt} showed that instances of self-admitted TD increase over the lifetime of a project and that it usually takes a long time for them to be resolved. \cite{Digkas2020CanDensity} and \cite{deSousa2020StudyingProject} did however also find that while the total amount of TD grows, the density decreases when new features with a lower TD density are added. \cite{deSousa2020StudyingProject} also found, through practitioner interviews, that inexperienced developers propagate error handling anti-patterns by replicating existing practices in the code base.

Two studies by Rios et al.~\citep{Rios2020HearingDebt}~\citep{Rios2020TheBrazil} asked practitioners to identify the causes and effects of \emph{documentation debt} (DD). Non-adaptation of good practices was the third most mentioned cause of DD\@. Similarly, low documentation quality (outdated\slash non-existent\slash inadequate) was a commonly mentioned effect of DD\@. Unfortunately, the studies do not make it clear how the practitioners might have interpreted the survey and interview questions (and some answers suggest that interpretations varied). Therefore, we cannot be sure whether the mentioned effects and causes refer to the specific instance of debt or the system in general. Only the latter interpretation would support the BWT\@.

\cite{Borowa2021TheDebt} investigated how bias contributed to \emph{architectural technical debt} (ATD) and found through interviews that biases such as the band-wagon effect, where one is likely to \say{do what others do}, influence architectural decisions.
\cite{Verdecchia2020ArchitecturalTheory} also interviewed practitioners and identified perceived effects of ATD\@. They found that ATD is perceived to cause code smells.
    
\cite{Besker2018EmbracingPerspective} studied TD management from the perspective of startups, through interaction with practitioners. They briefly mention a possible vicious cycle involving TD and project culture, but found that some developers saw TD not to be much of an issue in a static project group where everyone is aware of and used to it. They note a tendency to reduce TD before bringing new developers onto a project. In another study, \cite{Besker2019SoftwareWork} found that developers see themselves forced to introduce TD in as much as a quarter of the cases where they encounter existing TD; suggesting TD has a \emph{contagious} nature. \cite{Ozkaya2019TheDeveloper} also proposed a vicious cycle in which messy code reduces the morale of the developers, causing them to write more messy code.

Moreover, research has consistently shown a strong link between technical debt and vulnerability proneness in software. \cite{Izurieta2018Weaknessses} and \cite{Siavvas2022Enterprises} both found that technical debt indicators can serve as a potential indicator of security risks, with \cite{Izurieta2019Tactics} further emphasizing the importance of managing technical debt associated with cyber-security attacks.

Some studies have attempted to model the interest and growth of TD\@.
\cite{Martini2015TheCircles} conceptualizes a model detailing the cause and effect of ATD and find a potentially vicious cycle through which existing ATD causes more ATD to be implemented. Two other studies by Martini and Bosch~\citep{Martini2016AnAnaConDebt} \citep{Martini2017TheInterest} continue the work and evaluate methods in cooperation with industry for estimating the future cost of ATD\@. The derived methods involve several propagation factors of ATD, but none of the identified factors includes the BWT\@.
\cite{Martini2017OnPhenomenon} do however investigate how ATD propagates and one of their identified propagation patterns is \say{Propagation by bad example,} indicating that developers copy the practices they see in a system.

The literature search did not result in a single paper that treats the BWT as the main subject, and it is never discussed using the \emph{broken windows} terminology. The most common way it appears is by being brought up by interviewees as a cause or effect of TD\@. One of the most interesting entries is that by \cite{Tufano2015WhenBad}, which found that new code smells tend to appear in areas where code smells were already common. However, there are other possible explanations than BWT effects for that finding.

The concept of TD inducing further TD is not uncommon. However, the sub-mechanisms identified are usually restricted to that building on top of TD creates further TD, or that interaction effects between TD items result in higher interest payments than would have been necessary if the items were separate. The lack of studies on the possible psychological mechanics of how TD spreads is concerning as it may be one of the primary mechanics of runaway TD\@. Properly understanding these underlying mechanics of TD, allows us to refine existing models for how TD and its costs behave, helping practitioners establish effective management strategies that target these specific mechanics, and providing guidance for what future research might be warranted around TD and developer behavior.

We found no studies using an experimental methodology; most of them are field studies or judgment studies, making causality difficult to ascertain. One may conclude that this focus on more exploratory methods is a symptom of the research field being relatively new; some of the included entries explicitly state that the field has, until recently, chiefly been interested in definitions and theory-building. Furthermore, a majority of related literature has been published in the last five years.

\section{Method}
\label{sec:method}
\cite{Seaman1999QualitativeEngineering} argues that diverse evidence gathered through multiple distinct methodologies will constitute stronger proof than evidence of a single type, a concept commonly referred to as \emph{triangulation}. Following her advice, this study uses a mixed-methods design. The quantitative part consisted of a controlled experiment coupled with surveys, and the qualitative component was comprised of follow-up interviews. While more effort was put towards the former, the latter could constitute an essential complement, enabling us to clarify, contextualize and corroborate the results of the quantitative analysis.

\begin{figure}
    \centering
    \includegraphics[width=0.8\textwidth]{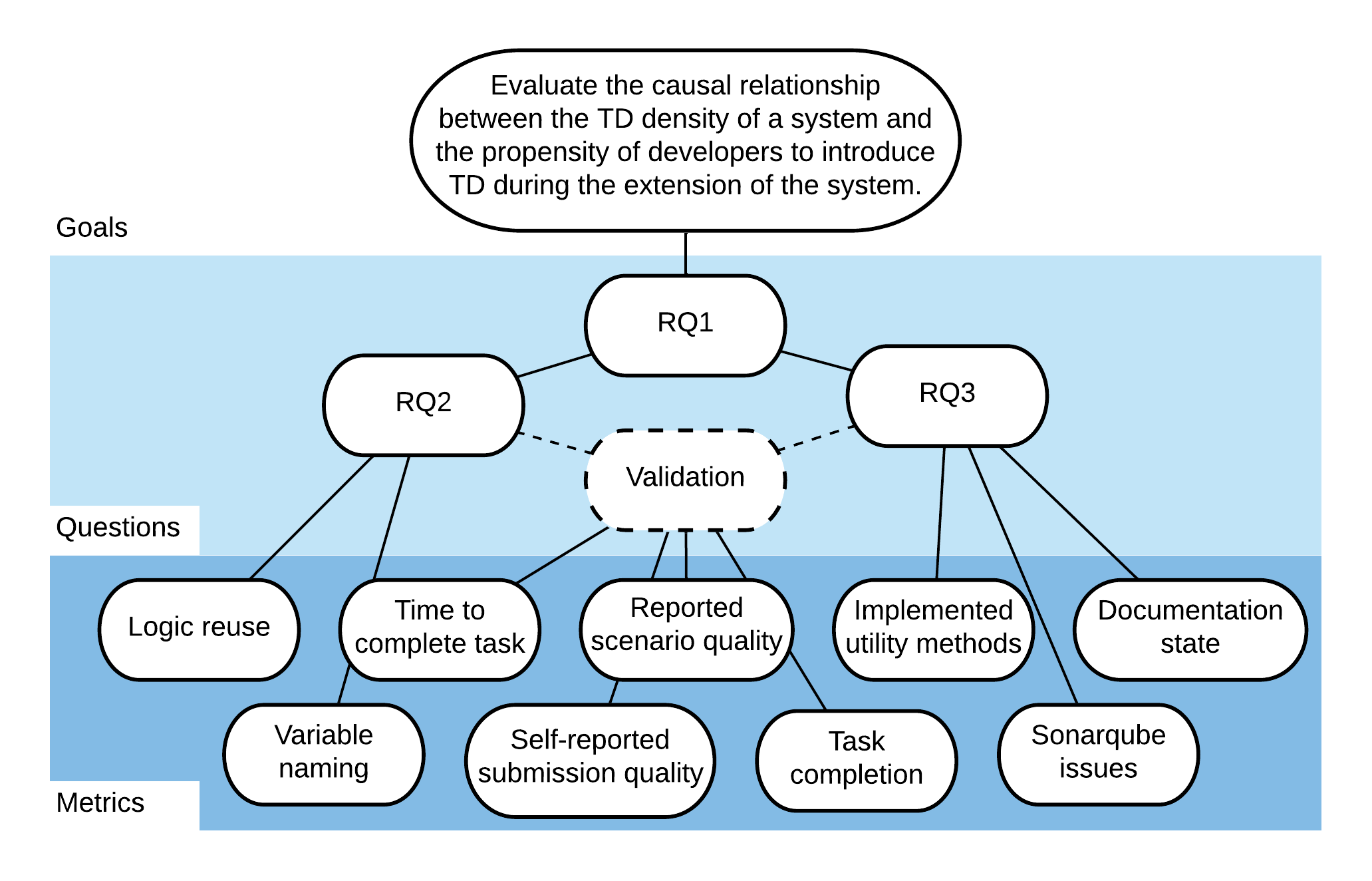}
    \caption[Goal-Question-Metric breakdown of the research objective.]{The GQM (goal, question, metric) breaks down the overarching goal (conceptual level) into questions (operational level) that needs answering in order to satisfy the goals. Answering the question requires identification and measurement of suitable metrics (quantitative level). Qualitative data from interviews were used to further validate and provide insight into our findings.}
    \label{fig:gqm}
\end{figure}

Using the software research strategy framework laid out by \cite{Stol2018TheResearch}, we would classify the experiment as \emph{experimental simulation} while we would argue that the interview component belongs in the \emph{judgment study} category. We consider them suitable complements to each other, as the former allows inference of \emph{causality}, while the latter does not. However, a judgment study is of a more flexible and exploratory nature, allowing us to discover some finer details that the experiment might miss~\citep{Stol2018TheResearch}. This section describes our methods for each component separately. Figure~\ref{fig:gqm} presents a Goal-Question-Metric breakdown of our design, as prescribed by~\cite{Basili2000TheParadigm}.%

\subsection{The Experiment}
\label{sec:experiment_design}

This section describes the design and execution of the experiment and survey that constitute the quantitative component of our study. The purpose of the experiment was to isolate the effects of the independent variable, \emph{technical debt} (TD), which would allow us to investigate the causal relationship.

To run these experiments we used a custom-built web-interface which meant that our only contact with the participants was recruitment and follow-up interviews. This was facilitated by providing participants with a sign-up code and information to access our research tool called \emph{RobotResearcher}. After receiving this information the participants could commence the experiment at a time suitable for them from the comfort of their browser.

\begin{figure}[ht!]
    \centering
    \includegraphics[width = \textwidth]{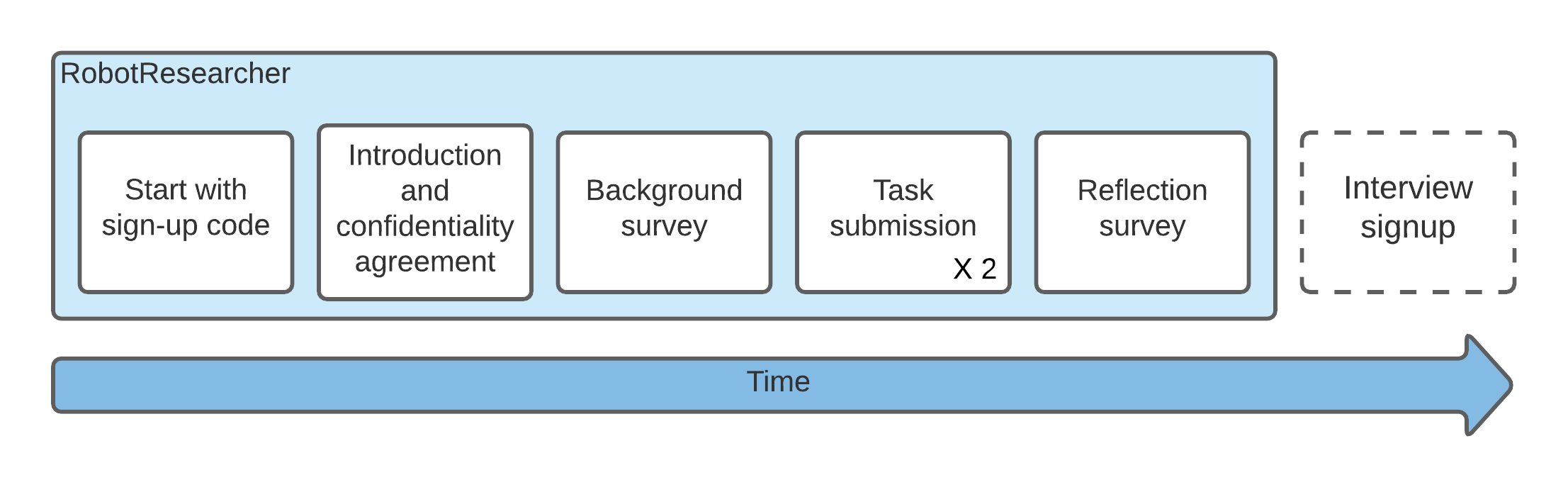}
    \caption[Experimental run overview from the perspective of a subject.]{Overview of a participant's experience with the data collection, showing the steps they had to complete. The step with a dashed border was optional.}
    \label{fig:data_collection}
\end{figure}

The experiment was split into several sequential steps as shown in Fig.~\ref{fig:data_collection}. After accessing the research tool, participants were presented with a short introduction about the experiment, as well as, a confidentiality agreement which they had to accept to proceed with the data collection part of the experiment. Next, participants were asked to complete a background survey to collect the data described in Sect.~\ref{sec:predictor_variables}. After the background survey, participants were taken to the main part of the experiment where they were presented with two tasks as described in Sect.~\ref{sec:treatment_scheme}. As a last step in the experiment, participants answered some reflection questions where they were asked to grade the quality of the preexisting code and their solution to both tasks. After completing the experiment all subjects were thanked for their participation and presented with a link to sign up for follow-up interviews.

\subsubsection{Technical Debt Level: The Independent Variable}
\label{sec:td_variable}
Since there is a vast, arguably infinite number of issues that constitute TD, determining a suitable subset to represent \emph{high TD} is not a trivial task.
The approach chosen was to go with \emph{two} distinctly different types of issues introduced at a very high rate, i.e. they were introduced everywhere where they could be introduced. This way, we ensured that designing suitable sets of systems and tasks, henceforth referred to as \emph{scenarios}, was feasible. It also mitigated the risk associated with choosing a single issue type as a representative for TD\@. While this choice won't allow us to conclude what effect every type of TD has it will allow us to draw conclusions about effects for at least \emph{some} types of TD\@.

The first type of TD we chose was \emph{bad variable names}, an issue with multiple great properties for the purposes of this study. Variable names are typically \emph{abundant} (in the vast majority of programming languages); hence changing them for the worse would constitute a flaw that the subject would be likely to detect. Bad variable names are also a clear example of \emph{code technical debt}. A variable whose name poorly describes its purpose will make the code less readable, which means that any time a piece of code needs to be understood, e.g., during refactoring, the process will be conceptually harder~\citep{Avidan2017EffectsStudy,Butler2009RelatingStudy}, and hence, slower. This extra time (and associated development costs) constitute the \emph{interest} payment that characterizes TD\@. An example of how bad variable names can make understanding harder is the difference between \texttt{a/b} and \texttt{distance/time} where it's a lot easier to deduce that the \texttt{velocity} was calculated in the second example.

The second type we chose was \emph{code duplication}. This issue shares the property of applying to the vast majority of programming languages; however, in other regards, it contrasts nicely with bad variable names. Notably, it may not be quite as obvious a defect, and avoiding propagation of it will usually take the developer more effort. In terms of scenario design, we found that accommodating the inclusion of code duplication and constructing tasks that a subject could solve with or without introducing additional TD were both feasible. Code duplication is often, but not always, an example of TD\@. Insisting on shared functionality between distant and unrelated classes may lead to excessive interdependency and complexity~\citep{Kapser2008cloningSoftware}. In many cases, however, there is a strong case that code duplication causes maintainability problems~\citep{Yu2012UnderstandingSystems} and thus can be considered TD~\citep{Alves2016}. 

Classifying code duplication as a particular type of TD is not straightforward. While~\cite{Li2015AManagement} firmly categorize it as code TD, using the taxonomy offered by~\cite{Alves2016}, it could be considered design TD\@. Further complicating matters, in a systematic review by \cite{Besker2018ManagingReview}, they found that several authors described code duplication as architectural TD\@. For the purpose of this study the exact classification is not important; we are content with noting that they all agree that code duplication often constitutes TD\@.

The above TD items could be introduced to the experiment scenarios without changing the overall structure of the code given to the participants. This made sure that most solutions to the tasks would be feasible in both the high and low-debt versions of the code base without making extensive changes to the existing code. This was essential as it allowed the participants to choose a solution, rather than being forced into one out of technical necessity.

\subsubsection{Controlled Variables}
When conducting an experiment, it is preferable to control as many factors as possible to avoid confounding. This study managed to control the following variables.

\textbf{Programming Language.} For several practical reasons, \textsf{Java}\footnote{\url{https://www.oracle.com/java/}} was chosen as the sole language of all scenarios. First, it is one of the most widespread programming languages in existence, not just in industry, but it is also often the language used for teaching object-oriented programming at educational institutions. Second, we, the authors, were all comfortable with working in \textsf{Java}.

\textbf{Development Environment.} To mitigate interference from participants using different editors with different levels of support we developed an original research tool that enforced a standard, but fairly bare-bones, editor. The editor is a web-based environment that integrates the ACE text editor\footnote{\url{https://web.archive.org/web/20210509090040/https://ace.c9.io/}} to provide a set of text editing shortcuts and syntax highlighting. The tool has access to a remote server that compiles and executes the users' code in a secure environment. It also oversees the experiment by allocating scenarios, treatments and providing instructions. A limited demo of the tool will remain available for evaluation purposes.\footnote{Demo available at \rrdemo{}, sources at \rrgithub{} and \rrdoi{}.}

\textbf{Scenarios} We created two tasks and their corresponding systems to be quite similar. One of them was to extend a room booking system with a new room type and the other was a commuting ticket system with a new ticket type. The scenarios were created such that they contained no or few technical debt items apart from those purposefully introduced. Both systems were of similar size, as well as structure, and are publicly available for review.\footnote{Availible at \scenariogithub{} and \scenariodoi{}}

\subsubsection{Dependent Variables}
The dependent variables represent what outcomes could be affected by the variation of the independent variable. We used various measures of the amount of TD added to the system by the subject. These particular measures were chosen as they were easily measurable and likely to appear in the limited context of the experiment tasks.

\textbf{Logic reuse} measured if the existing logic, validation logic and constructor, had been duplicated or reused while completing the task. For the purpose of this measure, any type of reuse qualified.

\textbf{Variable naming} recorded how many of the old, new, and copied variable names that fulfilled the following requirements, inspired by the guidelines proposed by \cite{Butler2009RelatingStudy}:

\begin{itemize}
    \item It should consist of one, or multiple, full words.
    \item Its name should have some connection to its use and\slash or the concept it represents.
    \item It should follow the format of some widely used naming conventions. While camel case is generally the standard for \textsf{Java}, we do not specify a standard and thus choose to generously also accept names that consistently follow other conventions such as snake case or pascal case.
    \item If it is used for iterating (where, e.g., \texttt{i,j} are commonly chosen names), or the variable represents something unknown (such as in the case of \texttt{equals()}) anything gets a pass.
\end{itemize}

\textbf{SonarQube issues} measured how many issues the static analysis tool \textsf{SonarQube} found in the submitted solution. Beyond the base set of issue identification rules (\textsf{Java Code Quality and Security}), we included three additional libraries \textsf{Checkstyle}\footnote{Checkstyle available at:  \url{https://github.com/checkstyle/sonar-checkstyle}}, \textsf{Codehawk Java}\footnote{CodeHawk available at:  \url{https://github.com/SEPMLAB/CodeHawk}}, and \textsf{Findbugs}\footnote{Findbugs available at:  \url{https://github.com/spotbugs/sonar-findbugs}}. The result is a set of 1,588 rules. Not all rules were suitable.
The following criteria were grounds for rule exclusion:\footnote{Full list of \textsf{SonarQube} rules used and excluded is available at \datagithub{} and \datadoi{}.}

\begin{itemize}
    \item The rule duplicates another rule, which was occasionally the case, as we included multiple libraries.
    \item The rule represented a \emph{possible standard}, i.e., one that may or may not be the standard at a particular institution but could not be considered part of general \textsf{Java} convention.
    \item The rule measures one of the things that we had already measured manually, e.g., code duplication detection rules were excluded in favor of manual inspection as we found them to be insufficiently effective.
    \item The rule concerns \emph{minor} indentation and cosmetic formatting errors. Such rules would generate copious amounts of issues, drowning out all others. We also considered them invalid since most commonly used development environments will help the user maintain proper indentation, while the provided editor would not.
    \item The rule did not make sense for a system as simple as the scenario code bases.
    \item The rule appeared not to function as intended.
\end{itemize}

\textbf{Implemented utility methods} recorded if the participant had implemented the utility methods \texttt{equals} and \texttt{hashCode} for the new classes they added. 

\textbf{Documentation state} measured whether the newly added functionality had been documented and whether the documentation was correct or not.

\textbf{Task completion} measured to which degree the task was completed on the scale: \say{Not submitted}, \say{Does not compile}, \say{Invalid solution}, and \say{Completed}.

\textbf{Time to complete task} was recorded as it could provide extra insight into the result of the study.

\subsubsection{Predictor Variables}
\label{sec:predictor_variables}
What we could not control, we accounted for through the inclusion of additional predictor variables. Some otherwise possibly confounding factors could be made part of the model by gathering background information on the subjects. 

The predictor variables were collected through a pre-task survey and included information about programming experience, work domain, education level and field. It also had questions regarding which of the following software practices were used at their most recent place of employment: \say{technical debt tracking}, \say{pair programming}, \say{established code standards}, and \say{peer code reviewing}.

\subsubsection{Treatment and Scenario Allocation Scheme}
\label{sec:treatment_scheme}
In anticipation of a modest sample size, we opted for a repeated measures design, i.e., a design that would entail each subject finishing multiple tasks. The drawback of such a design is that it can produce \emph{learning effects}, e.g., a subject may get better at a task with repetition. We took several measures to mitigate the impact of such effects:

\begin{itemize}
    \item The order of the scenarios and the treatment of scenarios was randomized.
    \item Each subject completed tasks in systems of high \emph{and} low levels of TD; this allowed the statistical model to isolate the individual `baseline' from the effect of the treatment.
    \item A subject was never given the same scenario twice. While the scenarios were similar enough to justify a fair comparison they were also different enough that a solution to one of them couldn't be easily adapted to the other one. Being able to reuse a previously created solution would likely have been irresistibly convenient.
\end{itemize}

The minimum number of tasks required to fulfill these criteria was two, each with an accompanying system that existed in a \emph{high TD} and a \emph{low TD} version. Hence, there would be four experimental run configurations in total, as shown in Table~\ref{tab:treatment_schedule}.

\begin{table}[ht]
    \centering
    \caption{Treatment Schedule.}
    \rowcolors{1}{lightgray}{white}
    \begin{tabulary}{\textwidth}{R|L}
    Scenario 1 - High TD & Scenario 2 - Low TD \\
    Scenario 1 - Low TD & Scenario 2 - High TD \\
    Scenario 2 - High TD & Scenario 1 - Low TD \\
    Scenario 2 - Low TD & Scenario 1 - High TD
    \end{tabulary}
    \label{tab:treatment_schedule}
\end{table}

\subsubsection{Study Validation}
\label{sec:validating_the_study}

Through a follow-up survey, we also asked the participants to rate the quality of their submissions as well as the system they were asked to work on as this data could help us validate the scenarios and how the quality of submissions was measured.

All classification and data extraction were done by two researchers independently and later compared to find and resolve any discrepancies or errors. All data from the experiment is publicly available.\footnote{Available at \datagithub{} and \datadoi{}}

As we used an original tool to collect data for the experiment we performed extensive user testing before the study was sent out. Both scenarios were in a similar manner subject to such tests to ensure that they were understandable and clear. These tests led to several improvements and were rerun until the testers reported a frictionless experience with the data collection tool and the scenarios themselves. No data from this testing was included in the study as the environment went through multiple changes and many of the testing participants had extensive knowledge about our hypothesis.

We also performed one pre-study run, and as we did not encounter any issues that led to revisions in our scenarios or data collection tool, we used all data from the pre-study in our analysis.

\subsubsection{Causal Analysis}
\label{sec:causal_analysis}
\label{background:causal_analysis}

\begin{figure}
    \centering
    \includegraphics[width=.7\textwidth]{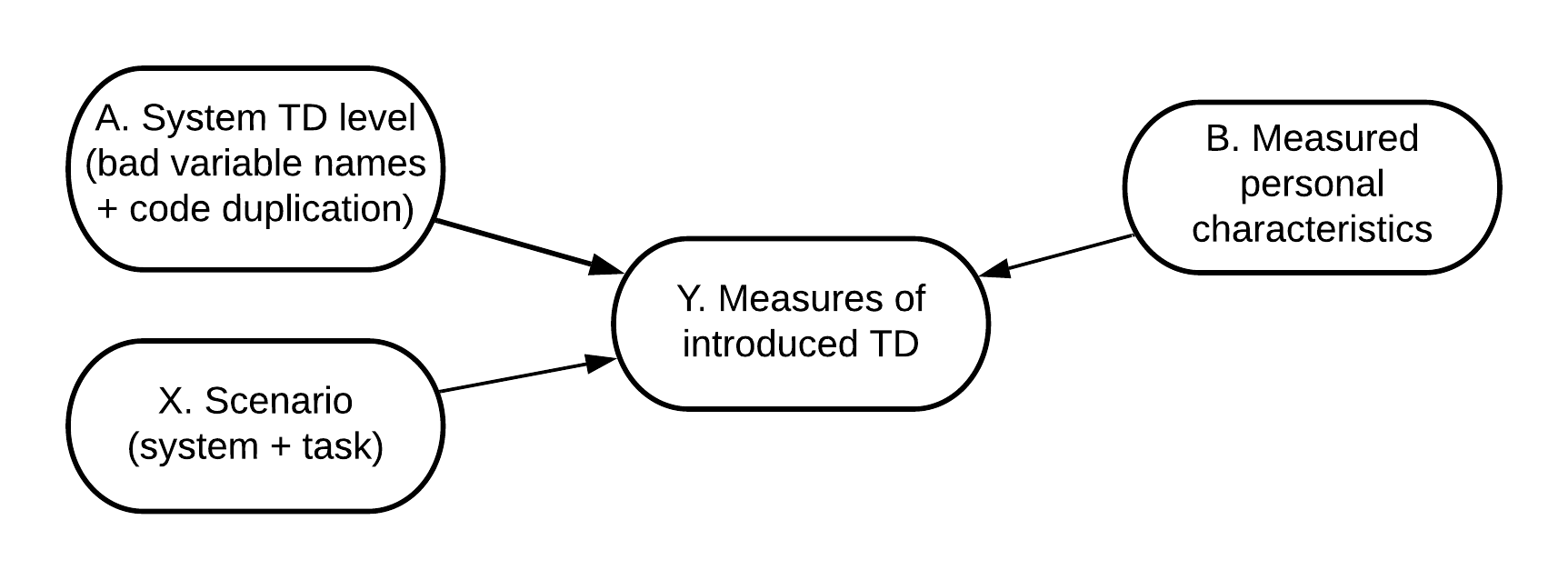}
    \caption[Directed acyclic graph encoding our causal assumptions concerning the experiment.]{Directed acyclic graph encoding our causal assumptions concerning the experiment. \say{Y. Measures of introduced TD} corresponds to \emph{logic reuse, variable, naming, \textsf{SonarQube} issues, documentation state, and implemented utility logic.} They are actually parallel (having the exact same ingoing and outgoing edges), but were condensed into a single node to keep the graph readable. The same goes for \say{B. Measured personal characteristics}, which combines all the measures taken during the background survey.} 
    \label{fig:causality}
\end{figure}

\cite{Pearl2011Causality:Edition} suggests building a \emph{directed~acyclic~graph}~(DAG) of all possible causal relationships in the analysis as it helps the authors document their assumptions and reason about, as well as make claims about causality. The DAG shown in Fig.~\ref{fig:causality} depicts the possible causal relationships between the measures we aim to safely include in our statistical models when we want to be able to infer causality. The causal relations of the DAG can easily be motivated. Both $A$ and $X$ were randomized and thus \textit{cannot} have been affected by any other variables. The personal characteristics ($B$) are measured before the participant is assigned a scenario and a TD level, hence protecting it from any influence of $A$ and $X$\@. The task submissions, from which the measures of introduced TD ($Y$) were derived, did not exist before the subject started the experiment, but were products of the participants, the scenarios, and the TD levels. Hence, they could \textit{not} have influenced the subjects' personal characteristics.

Once the DAG has been constructed, \emph{do-calculus} can be used to check if a specific causal relationship may be inferred from the statistical model~\citep{Pearl2011Causality:Edition}. This is done by constructing a \emph{do-statement} for the question we want to ask our model and transforming it using the \emph{rules of do-calculus} with the goal of eliminating all \emph{do}-clauses~\citep{Pearl2011Causality:Edition}. If all do-clauses can be removed it follows that a causal relationship can be measured by our statistical model~\citep{Pearl2011Causality:Edition}.

The questions we want to ask our model can be described as $P(\mathrm{Y}|\mathrm{do(X)}, \mathrm{A}, \mathrm{B})$. The second rule of do-calculus states that $P(\mathrm{Y}|\mathrm{do(X)}, \mathrm{Z})$ can be rewritten as $P(\mathrm{Y}|\mathrm{X}, \mathrm{Z})$ when all backdoor-paths between $\mathrm{X}$ and $\mathrm{Y}$ are blocked by $\mathrm{Z}$~\citep{Pearl2011Causality:Edition}. Our expression satisfies these criteria as $\mathrm{X}$ and $\mathrm{Y}$ are independent apart from the direct causal link $\mathrm{X}\rightarrow\mathrm{Y}$,  and when the rule is applied we receive $P(\mathrm{Y}|\mathrm{X}, \mathrm{A}, \mathrm{B})$ which tells us that a causal relationship can be measured by our model, assuming that the stated assumptions are correct. 

We will end this section by pointing out that the measured personal characteristics are \emph{not} randomized and that we can not, and will not, claim causality of any effects estimated for them. They could have numerous unmeasured ancestor variables in common with the dependent variable, creating causal backdoors that we can't block.

\subsubsection{Analysis Procedure}
\label{analysis:procedure}
\label{sec:BDA}

We used Bayesian data analysis (BDA) to analyze the data from our experiment. 
Bayesian data analysis allows the specification of a model, which is then fitted using empirical data through Bayesian updating~\citep{McElreath2020StatisticalRethinking}. This model can then be queried for multiple scenarios~\citep{McElreath2020StatisticalRethinking}. In contrast, a traditional frequentist analysis arrives at a result that is either \emph{significant} or \emph{insignificant}, which is entirely dependent on their selection of $\alpha$ (which commonly is set to $0.05$ in the natural sciences). Wasserstein et al.\ postulate that the fixation with \emph{statistical significance} is unhealthy, and thus, they encourage a move \say{beyond statistical significance}~\citep{Wasserstein2019Moving0.05}. Employing BDA is one way of following this suggestion which, according to Wasserstein et al., could prevent the software engineering community from ending up in a replication crisis like certain other disciplines.

A challenge with using BDA is that there is no widely accepted standard procedure. We largely followed the \emph{Bayesian workflow} laid out by~\cite{Gelman2020BayesianWorkflow}. However, there are many fine points to the iterative process of model building that they do not cover in detail.

We have multiple questions we would like to answer, each requiring its own specifically tailored model. To keep our analysis consistent, we used a uniform procedure to create and evaluate all models. Our complete analysis, conducted in accordance with the procedure detailed in this section, can be found in our replication package.\footnote{Available at \rppresent{}, source available at \rpgithub{} and \rpdoi{}}

We began the model building process by plotting the data and extracting some descriptive statistics (e.g., mean, variance, and median) to get a rough understanding of its distribution. We then created an initial model by choosing an appropriate distribution type and adding our essential predictors. We based this choice on whether the distribution type accurately described the underlying data generation process that produced the data and how well it fitted the empirical data. 

After adopting a distribution and basic predictors, we also had to set the prior probability distributions (priors) which encode our prior knowledge and belief into the statistical model. We did this by using the default priors (suggested by the \texttt{brm} function of the \textsf{brms} package\footnote{We used version 2.15.1, \textsf{brms} can be found at \textit{\url{https://github.com/paul-buerkner/brms}}}) as a starting point, and then tuned them until we had priors that allowed for physically possible outcome values. We also made sure that our $\beta$ parameters had priors that were skeptical of extreme effects. We did not encode any prior knowledge regarding the BWT into our priors since there appears to be no previous research on the subject in a software engineering context.

When we had an initial model with appropriate priors, we had to determine which predictors to include in the final model. For the models describing the amount of introduced TD, where we wanted to infer causality, we did this by creating a set of possible predictors subject to the restrictions laid out in Sect.~\ref{sec:causal_analysis}. For other models, we used a more exhaustive set of predictors.

Next, we created a new model for each of our possible additional predictors and compared the extended models with each other and the initial model using leave-one-out cross-validation (specifically the version implemented by the \textsf{Stan} package \textsf{loo}\footnote{\url{https://web.archive.org/web/20200814224945/https://mc-stan.org/loo/}}). The comparison allowed us to ascertain which predictors had significant predictive capabilities. Those predictors were then combined into a new set of expanded models. We repeated the process until there were no more predictors to combine.

As a last step we compared some of our most promising models and picked the simplest set of predictors which did not significantly reduce the models' performance. This model was slightly modified to increase the validity of our analysis by, e.g., fitting it to all the available data and adding $\beta$ parameters that provided us with extra insight. We then chose to keep those modifications if they did not negatively impact the sampling process or significantly harm the model's out-of-sample prediction capability.

The final model was used to answer our research questions. This was done partially by inspecting the estimates for the $\beta$ parameters we were interested in and querying the posterior probability distribution of the model, with factors fixed at various levels, and comparing the results. This final step of querying the model is crucial as it takes the general uncertainty of the model into account and provides an easy way to access effect sizes as it moves everything to the scale of the original input data~\citep{Torkar2018ArguingAnalysis}.

\subsection{The Interviews}
\label{sec:method_interviews}
This section describes the procedure we used to conduct the interviews that constitute the qualitative part of this study; the purpose of which was to gain further insights relating to our research questions. We also describe the analysis approach through which we analyzed the resulting material.

All participants who agreed to a follow-up interview were contacted and scheduled within a couple of days. We scheduled each to a maximum of $30$ minutes, but the length could vary significantly depending on the participants' availability.

The interviews were semi-structured around a set of questions with the purpose of giving us further insight into why the participants acted as they did in the experiment. We aimed for the questions to be open-ended while gradually steering the interviews toward the central topic of the study to capture as many unprovoked reflections as possible from each participant before revealing our research question. Given that we had $30$ allocated minutes, we were able to allow interviewees to elaborate on each question and go on semi-related tangents before we had to steer them back toward the topic of the study by asking the next question. The questions were:

\begin{enumerate}
    \item Can you describe your solutions to the tasks in the experiment?
    \item Can you motivate why you choose those solutions?
    \item What was your experience of the preexisting code?
    \item How did the preexisting code affect you?
    \item Do you have any other comments regarding the BWT in software engineering? (This question followed a brief explanation of the topic of our study)
\end{enumerate}

Interviews were documented using \emph{field notes} as described by~\cite{Seaman1999QualitativeEngineering}, i.e., we noted down any interesting reflections made by the interviewed participants relating to our research questions. We omitted any identifying details and, in many cases, translated the reflections from Swedish to English. Some participants who did not participate in the follow-up interviews provided us with their reflections in text. We also included such reflection in the interview material.

\subsubsection{Interview Results Analysis}
\label{sec:method_interviews_analysis}
The resulting list of noteworthy statements was subjected to \emph{thematic analysis} using a process largely following the steps laid out by~\cite{Braun2006UsingPsychology}, with the exception that our base material did not consist of a complete transcription but rather the field notes as described in Sect.~\ref{sec:method_interviews}. This is a brief description of the stages of the process~\citep{Braun2006UsingPsychology}:

\begin{enumerate}
    \item Familiarize yourself with the data.
    \item Encode patterns in the data as \emph{codes}.
    \item Combine codes into overarching \emph{themes}.
    \item Check theme coherency and accuracy against the data, iterate themes until those criteria are fulfilled.
    \item Define the themes and their contribution to the understanding of the data.
\end{enumerate}

When constructing themes, we did employ a partially \emph{deductive} approach, where we assumed some of the themes from the start, i.e., \emph{software quality} and \emph{system extension quality} (in terms of maintainability), since these were the areas of interest to us. The resulting set of themes was discussed and agreed upon by two of the authors.

\subsection{Participant Recruitment}
\label{sec:participant_recruitment}

We used convenience sampling by recruiting volunteers from our personal networks and local software development companies. As our personal networks mostly consisted of developers with little professional experience, we also used purposive sampling by specifically recruiting participants with more professional experience. Snowball sampling was also utilized by encouraging participants to pass on the participation invite. The sample used for the interviews was a subset of the experiment subjects, who volunteered after completing their participation in the experiment.

\subsection{Sample Description}
\label{sec:sample_description}
In total, $29$~subjects submitted solutions to at least one of the two tasks, for a total of $51$~submissions, meaning that seven subjects only submitted one of the tasks. Additionally, $14$~potential participants entered the research tool but did not do any tasks. The convenience sampling procedure employed produced a sample with some noticeable skews that reflect the nature of our personal networks and our geographical location (Gothenburg, Sweden).

\begin{figure}
    \centering
    \includegraphics[width=.9\textwidth]{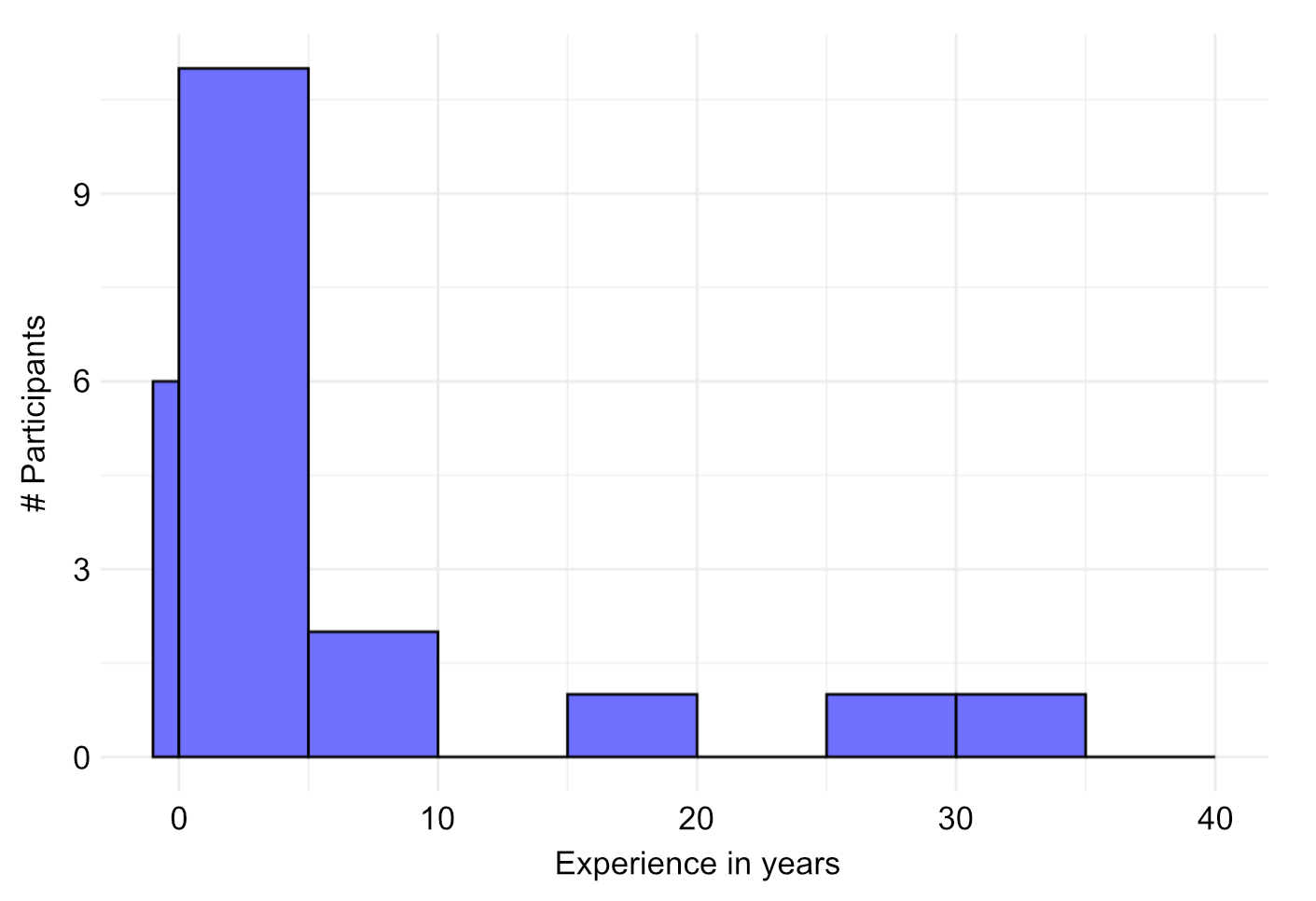}
    \caption[Professional experience of participants.]{Years of professional experience as reported by the participants.}
    \label{fig:experience}
\end{figure}

While the experience level of the participants varied, six of them had none or very little professional programming experience and an additional eleven participants had less than five years of experience. The full distribution of reported experience can be seen in Fig.~\ref{fig:experience}.

\begin{figure}
    \centering
    \includegraphics[width=.9\textwidth]{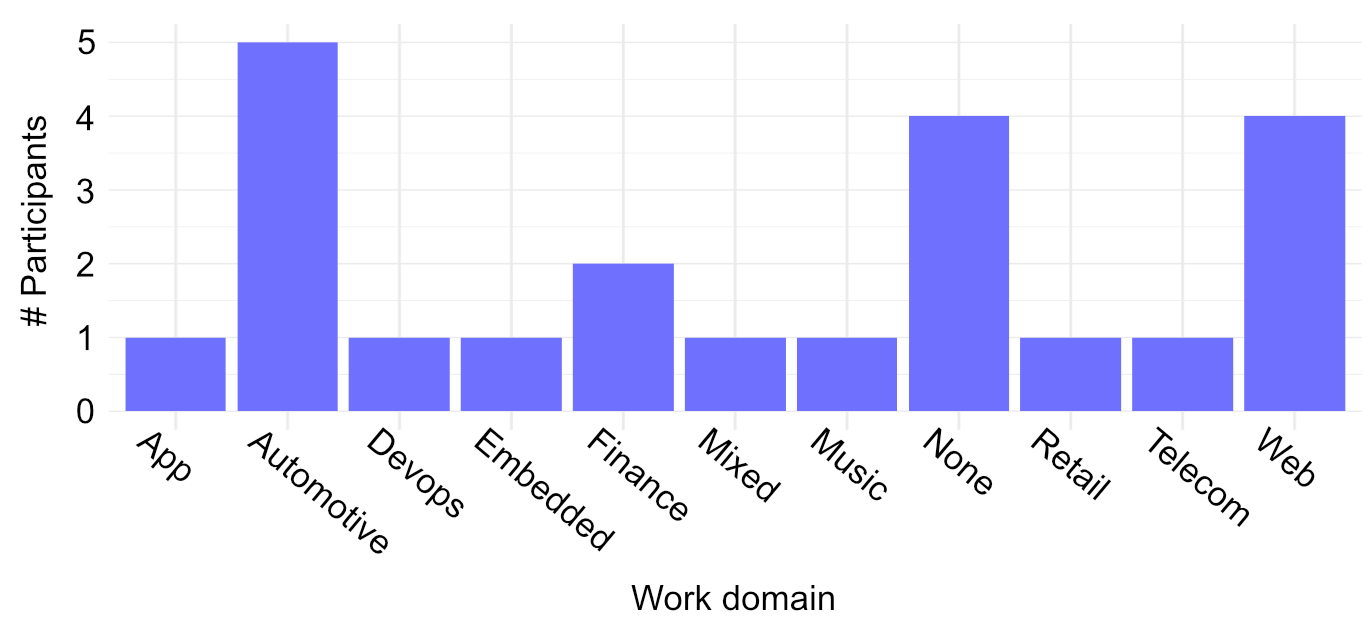}
    \caption[Software industry of participants.]{The software industry in which each of the participants reported having the most experience.}
    \label{fig:industry}
\end{figure}

As seen in Fig.~\ref{fig:industry} subjects reported working in a wide variety of software industry sectors ranging from embedded programming to web development. The \say{Automotive} section is over-represented among the participants, likely due to Volvo Cars being a large local employer.

A majority of participants reported having completed \say{some master-level studies,} with about a quarter claiming a finished master's degree. Regarding their field of study all but two participants reported \say{Computer Science} or \say{Software Engineering}, with the latter being twice as common as the former. Only one participant reported having no higher education.

A more detailed sample description with complementary descriptive statistics is available in the replication package.\footnote{Available at \rppresent{}, source available at \rpgithub{} and \rpdoi{}.}

\subsection{Method Deviations}
\label{sec:method_deviations}

Two participants reported that they had experienced problems with the user interface of the data collection tool. One of them accidentally reloaded the page and lost their progress. The other one accidentally submitted an empty solution, which lost us two submissions. However, these drop-outs could be considered \emph{random} and should not bias the sample.

One of the early participants discovered two minor mistakes in the given scenarios. We resolved\footnote{Diff showing resolution: \url{https://github.com/BWTSE/Scenarios/compare/mainstudy-v1...mainstudy-v2}} the mistakes upon notice and presented the revised versions to subsequent participants. Given the limited impact of the changes---they were not in a location that participants frequently modified---we find it unlikely that many participants noticed, let alone were impacted by their unfortunate presence.

\section{Quantitative Results}
\label{sec:quantitative}

This section presents the estimates and some carefully selected posterior samples produced by the final models for each outcome. Four outcome models describing how likely the developers were to implement utility methods, add documentation to their code, complete the task, and the time it took to complete the task, are not presented in detail in this section as they showed no significant effects. Furthermore, including all the intermediary models described in Sect.~\ref{analysis:procedure} would not be feasible. Instead, we have provided a replication package\footnote{Available at \rppresent{}, source available at \rpgithub{} and \rpdoi{}.} that allows thorough examination and reproduction of all our models and their development.

As described in Sect.~\ref{sec:sample_description}, seven participants failed to submit both tasks and therefore only contributed one submission each. These drop-outs were analyzed and were found to have no correlation with either treatment or scenario allocation. While multiple measures from each participant would have been desirable and lead to higher certainty in our models we can safely include all the submissions in our analysis while considering both within and between subjects effects as we use partial pooling, which will ensure that single data points do not skew the result.

\begin{table}[ht]
\centering
\caption[Population level effects]{Population level effects. The \texttt{intercept} estimates correspond to the baseline of the measurements (with all predictors set to zero, or \emph{true} in the case of Boolean values). Models of the \emph{cumulative} family, has multiple intercepts where \texttt{intercept[1]} is the logarithmic chance of observing the first outcome rather than the remaining. \texttt{intercept[2]} is the chance of observing one of the first two outcomes rather than the remaining, and so on. The $\beta$ estimates correspond to how much a parameter influences the outcome. Consult the replication package for the full distribution of the $\beta$ estimates.
}
\label{fig:pop-eff}
\begin{tabular}{llrrrr}
\textbf{Model} & \textbf{Parameter}                         & \textbf{mean} & \textbf{sd} & \textbf{l-95\% CI} & \textbf{u-95\% CI} \\ \hline
(1) Validation logic reuse  & \texttt{intercept}                         & $-0.16$       & $0.57$      & $-1.30$            & $0.96$  \\
&\texttt{$\beta$-high\_debt\_version:false} & $-1.79$       & $0.66$      & $-3.12$            & $-0.54$             \\
&\texttt{$\beta$-programming\_experience}   & $0.26$        & $0.47$      & $-0.67$            & $1.21$          \\ \hline
(2) Constructor logic reuse & \texttt{intercept}                         & $-0.38$       & $0.58$      & $-1.53$            & $0.74$             \\
&\texttt{$\beta$-high\_debt\_version:false} & $-1.63$       & $0.65$      & $-2.95$            & $-0.39$             \\
&\texttt{$\beta$-programming\_experience}   & $0.31$        & $0.48$      & $-0.66$            & $1.24$    \\ \hline
(3) Variable naming & \texttt{intercept}                         & $1.52$        & $0.50$      & $0.65$             & $2.62$             \\
&\texttt{$\beta$:high\_debt\_version:false} & $2.48$        & $0.57$      & $1.41$             & $3.64$             \\
&\texttt{$\beta$:programming\_experience}   & $0.14$        & $0.46$      & $-0.75$            & $1.09$  \\ \hline
(4) Introduced \\ \textsf{SonarQube} issues & \texttt{intercept}                          & $0.73$        & $0.30$      & $0.15$             & $1.32$             \\
&\texttt{$\beta$:high\_debt\_version:false} & $-0.80$        & $0.37$      & $-1.53$            & $-0.08$            \\
&\texttt{$\beta$:programming\_experience}   & $-0.23$        & $0.23$      & $-0.68$            & $0.23$     \\ \hline
(5) System Quality \\ Rating (cumulative) & \texttt{intercept[1]}   &   $-2.08$      &   $0.54$    &   $-3.21$    &   $-1.07$             \\
&\texttt{intercept[2]}   & $-0.59$      &   $0.40$    &   $-1.41$     &   $0.17$             \\
&\texttt{intercept[3]}    &   $0.30$      &   $0.40$    &   $-0.50$     &   $1.07$             \\
&\texttt{intercept[4]}   &   $0.81$      &   $0.41$     &   $0.01$     &   $1.62$             \\
&\texttt{intercept[5]}   &   $2.07$      &   $0.47$     &   $1.16$     &   $3.04$             \\
&\texttt{intercept[6]}   &   $4.21$      &   $0.74$     &   $2.92$     &   $5.77$             \\
&\texttt{$\beta$:high\_debt\_version:false} & $1.41$   &   $0.47$    &  $0.48$   &  $2.33$             \\
&\texttt{$\beta$:programming\_experience}   & $-0.52$      & $0.27$    & $-1.07$    & $-0.01$ \\ \hline
(6) Self-reported submission \\ quality (cumulative) & \texttt{intercept[1]}                         & $-4.81$        & $1.25$      & $-7.39$             & $-2.48$             \\
&\texttt{intercept[2]}                         & $-3.53$        & $1.07$      & $-5.71$             & $-1.45$             \\
&\texttt{intercept[3]}                         & $-1.46$        & $0.98$      & $-3.43$             & $0.45$             \\
&\texttt{intercept[4]}                         & $1.43$         & $0.99$      & $-0.48$             & $3.47$             \\
&\texttt{intercept[5]}                         & $2.17$         & $1.02$      & $0.20$              & $4.24$             \\
&\texttt{intercept[6]}                         & $3.12$         & $1.11$      & $0.99$              & $5.41$             \\
&\texttt{$\beta$:var\_naming\_copied:good}        & $0.19$        & $0.66$      & $-1.10$            & $1.52$             \\
&\texttt{$\beta$:var\_naming\_new:good}           & $-0.04$       & $0.77$      & $-1.53$            & $1.46$             \\
&\texttt{$\beta$:reused\_logic\_validation:false} & $-1.31$       & $0.66$      & $-2.62$            & $-0.03$             \\
&\texttt{$\beta$:equals\_exists:false}            & $-0.17$       & $0.56$      & $-1.25$            & $0.97$             \\
&\texttt{$\beta$:sonarqube\_issues}               & $0.09$        & $0.31$      & $-0.54$            & $0.68$                 \\
&\texttt{$\beta$:documentation:incorrect}         & $-0.19$        & $0.67$      & $-1.51$            & $1.12$                 \\
&\texttt{$\beta$:documentation_none}              & $-0.33$       & $0.61$      & $-1.53$            & $0.91$                 
\end{tabular}
\end{table}

\begin{tcolorbox}
\textbf{Interpreting the Results}

The \emph{predictors} discussed in this section are factors that showed some predictive power on an outcome. The magnitude of the influence of a predictor on the outcome is described by a \emph{$\beta$ (parameter) estimate}. In Bayesian data analysis, the $\beta$ estimate is not a single value but a probability density.

When interpreting the results, the mass of the $\beta$ distributions in relation to zero is important as it represents the likelihood of a parameter having an effect. However, to infer anything about the \emph{size} of that effect, the \emph{outcome distribution} is vital as it shows the effects on the same \emph{scale} as the data. This allows us to evaluate the \emph{practical} significance.
\end{tcolorbox}

All models include $\beta$ parameters for \texttt{high\_debt\_version} and \texttt{programming\_-\\experience} as well as a varying intercept for each participant (\texttt{session}). The $\beta$ parameter for \texttt{high\_debt\_version} is compulsory as examining the effects produced by the existence of TD is the point of our study. The parameter for \texttt{programming\_experience} was included as we want some insight into what effect the skew in our sample towards more junior developers might have on our result.
We did not consider any interaction effects and did not add any other grouping factors than the participant identifier since they would entail a high risk of overfitting with our relatively small sample size.

We fitted all final models to the full data set, including participants who only completed one of the assigned tasks. Excluding the participants who did not complete both tasks made sampling easier as it assured that the data set was balanced but could also introduce bias in the analysis. Therefore, we chose to fit the final models with all available data after confirming that the models could sample well using the complete data set.

\subsection{Logic Reuse}
\label{sec:result_reuse}

Table~\ref{fig:pop-eff} (Models 1 \& 2) show that the \texttt{high\_debt\_version} predictor has a significant effect on the outcome with well over $95\%$ of the $\beta$ estimate probability mass on the negative side of zero. The $\beta$ estimate distribution for the \texttt{programming\_experience} predictor is centered close to zero with significant weight on both sides, indicating no or minimal effect. This information alone indicates that a high debt density in the preexisting code base induces developers to reuse less. In contrast, the developers' professional programming experience scarcely affects the amount of reused code.

\begin{figure}
    \centering
    \includegraphics[width=.9\textwidth]{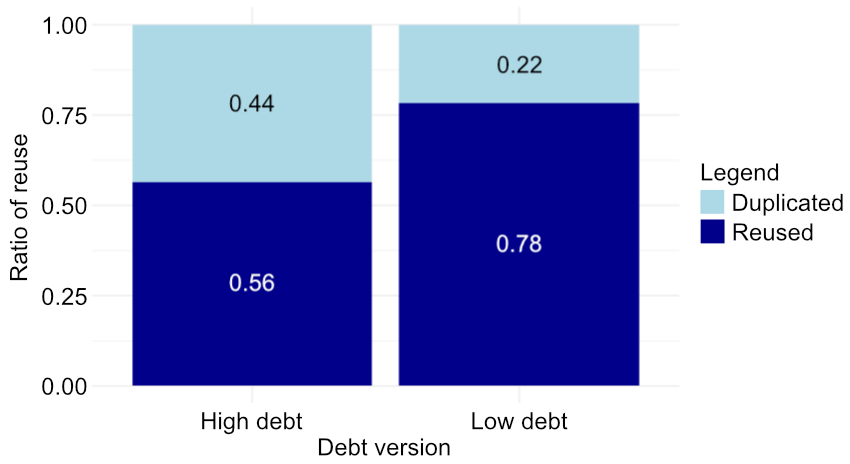}
    \caption[Outcome distribution of logic reuse.]{Outcome distribution of the model describing reuse, separated by high and low debt versions. The outcome was simulated for developers with $10$ years of professional programming experience.}
    \label{fig:reuse-posterior}
\end{figure}

The outcome distribution shown in Fig.~\ref{fig:reuse-posterior} offers additional insights into the effects of the predictors. It shows a clear difference in the outcome depending on the TD level of the system. The effect appears to persist for all levels of professional programming experience. The model estimates that developers with $10$~years of professional programming experience are $102\%$~more likely to duplicate logic in the high debt version of the system. The corresponding number for those with no professional programming experience is~$113\%$. Given that this model was developed in accordance with the restrictions arrived at in our \emph{causal analysis} (Sect.~\ref{sec:causal_analysis}), it is possible to make causal inferences regarding the effects of TD level (but \emph{not} experience level).

\subsection{Variable Naming}
\label{sec:result_var_names}

Table~\ref{fig:pop-eff} (Model 3) shows that the \texttt{high\_debt\_version} predictor has a significant effect on the outcome with well over $95\%$~of the probability mass on the right side of zero. Conversely, the \texttt{programming\_experience} predictor's $\beta$ estimate is centered around zero, indicating no or minimal effect. These results imply that a high debt density in the preexisting code base causes developers to use non-descriptive variable names and that the developers' professional programming experience has little to no effect on the descriptiveness of their variable names.

\begin{figure}
    \centering
    \includegraphics[width=.9\textwidth]{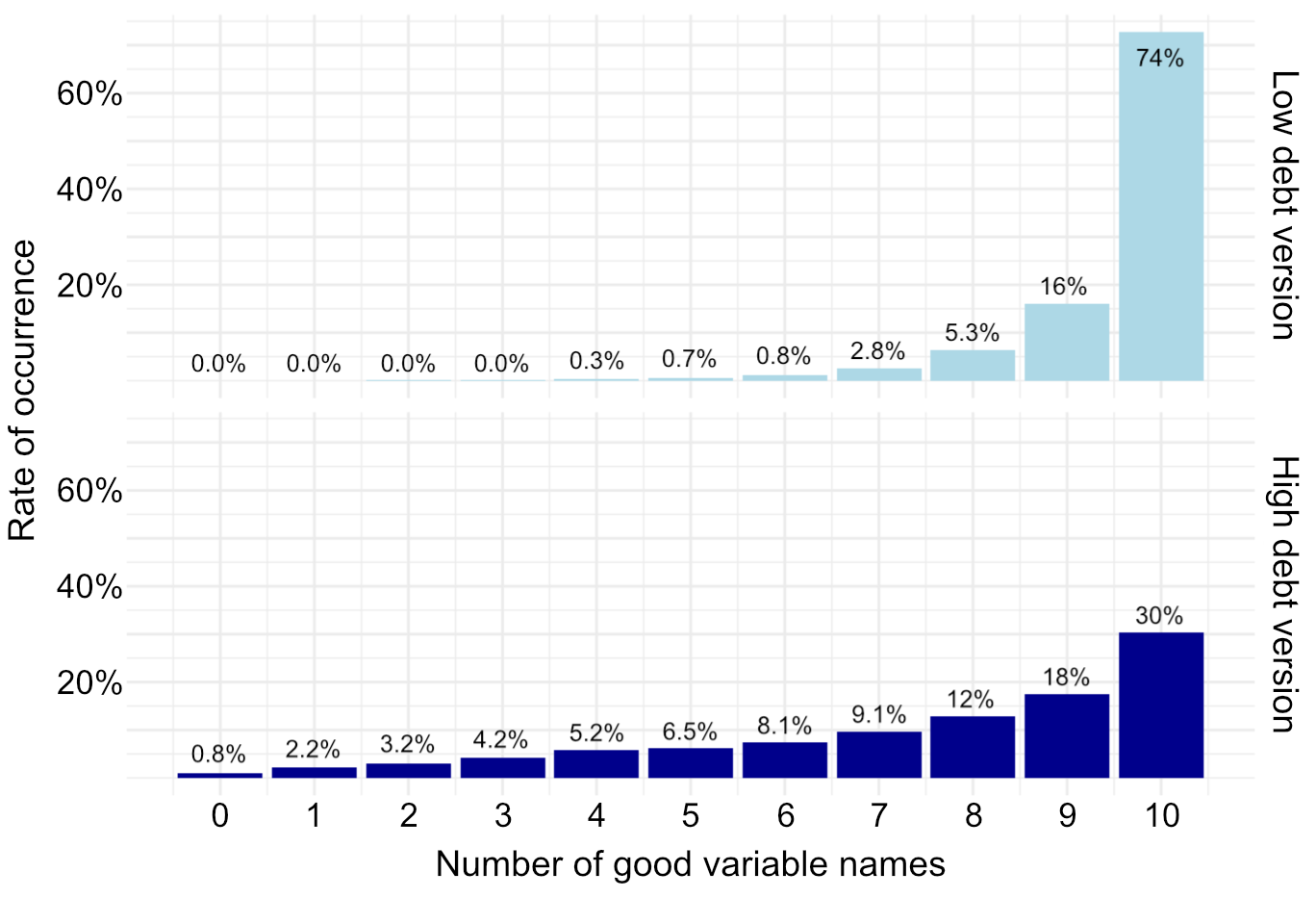}
    \caption[Outcome distribution of variable naming, ten variable example.]{Outcome distribution of the model describing variable naming. The histogram shows the estimated probability of each outcome (i.e., number of descriptive variable names) when a developer with ten years of professional experience introduces ten new variables}
    \label{fig:vars-posterior-real}
\end{figure}

Figure~\ref{fig:vars-posterior-real} shows the distribution of the estimated rate of good variable naming for a developer with ten years of professional programming experience introducing ten variables. According to the depicted sample, a developer is $458\%$~more likely to use a non-descriptive variable name whenever they introduce a variable in the presence of TD\@. Given that this model was developed in accordance with the restrictions arrived at in our \emph{causal analysis} (Sect.~\ref{sec:causal_analysis}), it is possible to make causal inferences regarding the effects of TD level (but \emph{not} experience level).

\subsection{SonarQube Issues}
\label{sec:result_sonar}

The $\beta$ estimates presented in Table~\ref{fig:pop-eff} (Model 4) suggest that the debt level of the system has a notable effect on the number of \textsf{SonarQube} issues introduced by the subject with the $95\%$~credible interval of the estimated \texttt{high\_debt\_version} parameter outside of zero. The \texttt{programming\_experience} $\beta$ estimate is centered closer to zero, but the uneven distribution suggests that it could potentially have some effect.

\begin{figure}
    \centering
    \includegraphics[width=.9\textwidth]{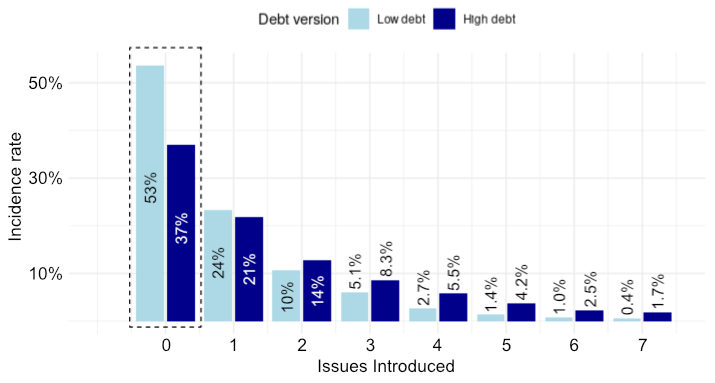}
    \caption[Outcome distribution of introduced \textsf{SonarQube} issues.]{Outcome distribution of the model describing the number of \textsf{SonarQube} issues, separated by high and low debt versions as well as years of professional programming experience.}
    \label{fig:sq-posterior}
\end{figure}

The outcome distributions depicted in Fig.~\ref{fig:sq-posterior} show a moderate effect of the debt density of the system on the number of \textsf{SonarQube} issues introduced by the participant. The, somewhat unreliable, estimated effect of professional programming experience generates noticeable differences in the prediction for various experience levels, with more experience being expected to introduce fewer \textsf{SonarQube} issues. The effects shown by the graph correspond to developers on average introducing $117\%$~more issues in the high debt version. Given that this model was developed in accordance with the restrictions arrived at in our \emph{causal analysis} (Sect.~\ref{sec:causal_analysis}), it is possible to make causal inferences regarding the effects of TD level (but \emph{not} experience level).

\subsection{System Quality Rating}
\label{sec:result_qual_rating}

As shown by Table~\ref{fig:pop-eff} (Model 5) this model estimates considerable effects of TD level (\texttt{high\_debt\_version}) and professional programming experience on the way subjects rated the the system in terms of quality (maintainability). In both cases, the $95\%$~credible intervals do not cross zero. 
These results show that participants tended to rate the \emph{high TD} as worse and more experienced programmers also, on average, give all systems a lower rating.

\begin{figure}[ht!]
    \centering
    \includegraphics[width=.9\textwidth]{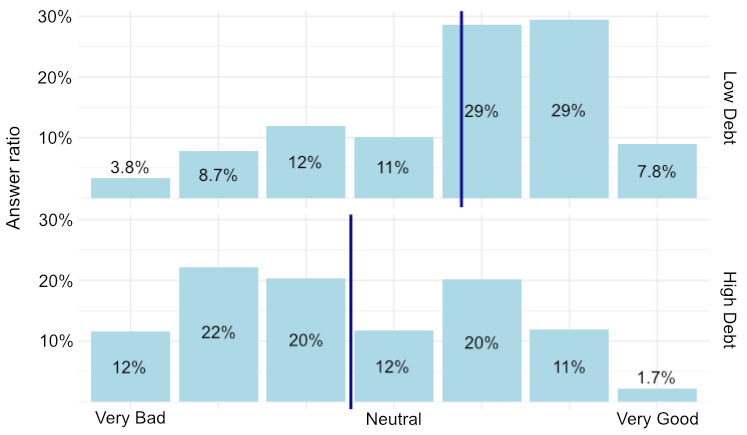}
    \caption[Outcome distribution of system rating, low experience.]{Outcome distribution of the model describing system rating in terms of quality (maintainability). Simulated for developers with $3$~years of programming experience, and separated by debt level of the system.} The vertical lines represent the means.
    \label{fig:rep-qual-low-exp}
\end{figure}

\begin{figure}[ht!]
    \centering
    \includegraphics[width=.9\textwidth]{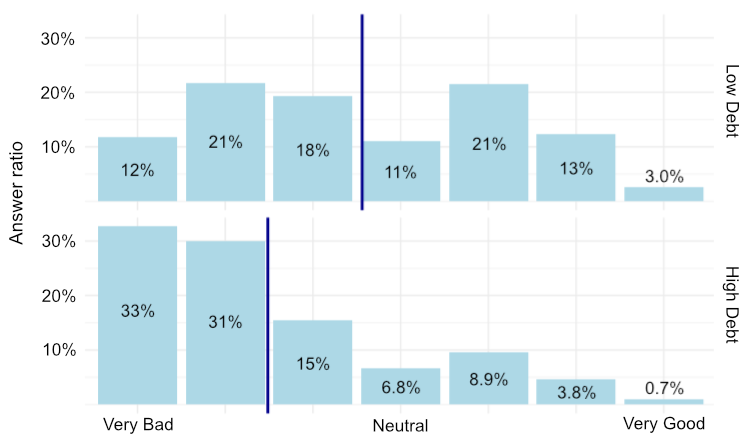}
    \caption[Outcome distribution of system rating, high experience.]{Outcome distribution of the model describing system rating in terms of quality (maintainability). Simulated for developers with $25$~years of professional programming experience and separated by debt level of the system.} The vertical lines represent the means.
    \label{fig:rep-qual-high-exp}
\end{figure}

The outcome distributions differ noticeably between the high and low debt cases in both our low experience (Fig.~\ref{fig:rep-qual-low-exp}) and high experience (Fig.~\ref{fig:rep-qual-high-exp}) cases. The average difference is about one unit on the Likert scale. The model also predicts that a developer with ten years of experience is $197\%$ more likely to rate the high debt system as worse than the low debt system, as opposed to the other way around.

\subsection{Self-Reported Submission Quality}
\label{sec:result_self_rating}

Table~\ref{fig:pop-eff} (Model 6) shows a variety of small effects with high uncertainty. However, the effect of \texttt{reused\_logic\_validation:false} is significant enough that the $95\%$~credible interval does not cross zero, indicating that the amount of duplication is strongly linked to how participant rated their own work.

\begin{figure}
    \centering
    \includegraphics[width=.9\textwidth]{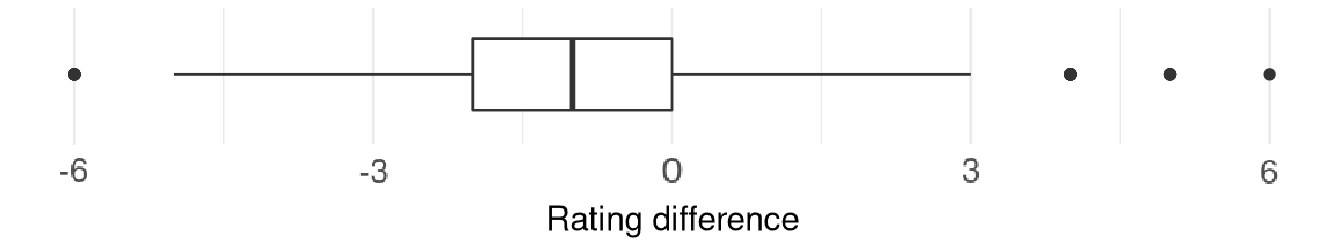}
    \caption[Distribution of differences in outcome of subjects' self-rating of their submissions.]{The distribution of differences between the participants' quality rating of their own work, simulated for a \emph{bad} submission with a lot of introduced TD and a \emph{good} submission with close to no introduced TD according to our measurements. The vertical lines represent the means. Difference as: bad submission rating $-$ good submission rating.}
    \label{fig:self-posterior-diff}
\end{figure}

Further insight is offered by Fig.~\ref{fig:self-posterior-diff}, which shows the distribution of differences between outcomes generated under the assumption of a \emph{bad} submission and \emph{good} submission. In the simulated case, the model estimates that participants are $114\%$~more likely to rate the bad submission as worse than they were to rate the good submission as worse.

\section{Qualitative Assessment}
\label{sec:assesment}
Our results concerning the introduction of further \emph{technical debt} (TD) in the form of logic duplication, bad naming of newly created variables, and other issues discovered through the use of \textsf{SonarQube} all showed considerable effects:

\begin{itemize}
    \item Developers are $102\%$~more likely to duplicate existing logic in our systems with high levels of TD\footnote{\label{finding_note_1}Effect estimated for a developer with $10$ years of professional programming experience that is working on the tasks and systems provided in this study} (Sect.~\ref{sec:result_reuse}).
    \item Developers are $458\%$~more likely to assign a variable a non-descriptive name in systems with high levels of TD\textsuperscript{\ref{finding_note_1}} (Sect.~\ref{sec:result_var_names}).
    \item Developers introduce $117\%$~more \textsf{SonarQube} issues in our systems with high levels of TD\textsuperscript{\ref{finding_note_1}} (Sect.~\ref{sec:result_sonar}).
\end{itemize}
 
All of these were established by investigating $95\%$ credible intervals. Taken together, it is very improbable that they are all false positives. Additionally, the results of our \emph{causal analysis} (see Sect.~\ref{sec:causal_analysis} show that we can infer the \emph{causality} of these relationships, i.e., pre-existing TD is the \emph{cause} of the effects.

\begin{tcolorbox}
\textbf{Finding~1 (RQ1}): Existing TD increases the likelihood of developers introducing new TD when extending a system, even in cases where it is not \emph{necessary} to do so.

\textbf{Finding~2 (RQ2}): Existing TD of a certain type increases the likelihood of developers introducing new TD of that type when extending a system, even in cases where it is not \emph{necessary} to do so. 
\end{tcolorbox}

We found no noticeable effect of TD on a subject's propensity to implement utility methods for the new classes they constructed. Similarly, we found no significant effect of pre-existing TD on the likelihood of participants \emph{correctly} documenting their new code. This does however not mean that we can reject the possibility of such relation as the probability density, while crossing zero, is quite wide showing a large uncertainty in these estimates.

The experiment design, i.e., bad variable names and code duplication coinciding, makes it impossible for us to discern the effects of the former from the latter. However, the estimated impact on the introduction of general code smells suggests that \emph{broken windows theory} effects are not limited to within specific types of TD items. That is, pre-existing bad variable names do not just induce the developer to implement additional poorly named variables, but also other examples of TD\@. This finding is interesting because it suggests that developers mimicking previous work is not a sufficient explanation of BWT effects on its own. 

\begin{tcolorbox}
\textbf{Finding 3 (RQ3}): Mimicry of existing instances of TD is not, alone, a sufficient explanation of BWT effects in software engineering.
\end{tcolorbox}

Furthermore, our analysis of the participants' evaluation of their work (Sect.~\ref{sec:result_self_rating}) reveals a tangible correlation between the way a subject rated their submitted changes in terms of quality (maintainability) and some of our various measures of TD\@. This correlation shows that subjects were not completely oblivious regarding the TD they had introduced.

\begin{tcolorbox}
\textbf{Finding 4:} Developers appear to be, at least partially, aware of their introduction of TD\@.
\end{tcolorbox}

\subsection{Thematic Analysis}
\label{sec:qualitative}
The data set forming the basis of our thematic analysis were field notes taken during six follow-up interviews as well as additional comments received via email or other text messages from a further four participants. This section presents the results of that analysis. 

\begin{figure}
\centering
\includegraphics[width=.8\textwidth]{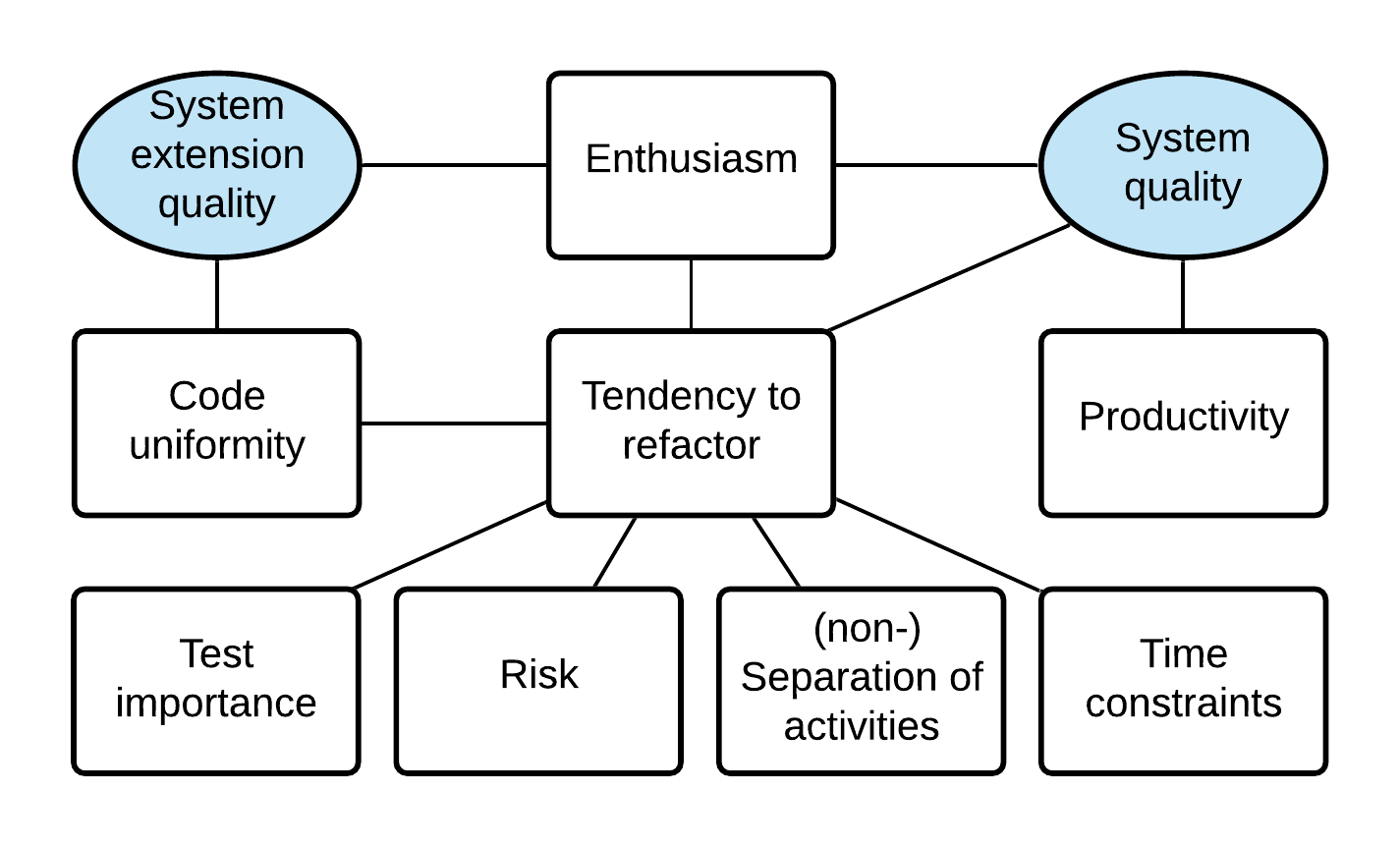}
\caption[Thematic map derived from interview results.]{A map of the themes produced by our thematic analysis. The oval nodes represent the themes that we had decided should be included before starting the thematic analysis. Nodes connected by vertices occurred together in the interview material.}
\label{fig:thematic_map}
\end{figure}

We arranged the themes derived through this process into the thematic map that constitutes Fig~\ref{fig:thematic_map}. A connection between two themes represents them appearing together in the interview material. The manner of their connection may differ between interviews.

The oval themes represent the ones that we enforced as part of our deductive approach. They are the focal points of our inquiry, the TD density or quality (maintenance) of the system, and the same properties of the new code they produce.

The subjects were clearly bothered by the issues in the high TD version. Some complained that the short and non-descriptive variable names made the system difficult to understand, which affected their productivity. While expected, it is a solid indication that bad variable names are a good example of TD\@. We also found that some expressed a sense of discouragement or loss of enthusiasm in relation to system quality exemplified by comments such as \say{It was such a drag with all the bad variable names \ldots//\ldots I almost didn't finish the task}. 

Refactoring was a topic frequently brought up and is very much a topic of interest in the context of TD propagation. The task descriptions explicitly stated that subjects were free to alter other parts of the system as they saw fit, and most of our interviewees had taken advantage of that opportunity. Interestingly several of them connected system quality to their decision to refactor, while some argued that refactoring a system that was in such a sorry state was not worth their time, others expressed that the obvious faults were what prompted them to start refactoring. One said that the quality issues caused him to go on a \say{refactoring spree} and another that the noticeable faults made them examine the system more closely, in search of more quality issues in need of attention.

Views on refactoring varied significantly, some older and more experienced developers favored an approach where refactoring efforts are mainly relegated to specifically dedicated time-slots: \say{You should devote specific sprints to refactoring and not mix it with feature development}. In their view, mixing refactoring work with new feature development results in productivity losses and results in individual developers exposing the project to the risks they associate with refactoring (instead, they favor team deliberations regarding significant refactoring decisions). Others favored a more flexible approach where they fix problems as they come across them, though admitting that \say{going down the rabbit hole} (having to follow a seemingly never-ending trail of interdependent issues) is a definite risk, especially when dealing with older systems whose architecture is reminiscent of a \say{ball of yarn}. Finally, high test quality and coverage were mentioned as a way of enabling more aggressive refactoring strategies by mitigating the associated risks.

Several participants noted that they had trouble deciding on how to name the variables in their new class when working on the high debt system. While they recognized that the current naming scheme was terrible, using a different one for their new variables would make the code less uniform, which could potentially be more confusing. One said that \say{it's important to follow the style of the code already there to reduce bugs.} The optimal solution, they agreed, would be to fix all variable names (which some actually did), although one expressed regret over not having done it. This participant went on to admit that their discouragement, as a result of the quality issues, had made them focus on just getting the tests to pass as soon as possible to \say{be done with it}. 

Our impression is that those who volunteered to be interviewed were more enthusiastic about the project, generally put more effort into their submissions than the average subject, and were excited to discuss the qualities of their particular solution.

\section{Discussion}

As summarized in Sect.~\ref{sec:assesment} our results clearly show that developers' propensity to introduce TD is higher while working in a system with more preexisting TD (RQ1). Moreover, we see the same effect when only looking at introduced TD of similar types (RQ2), as well as dissimilar types (RQ3) of TD compared to the preexisting TD\@. The thematic analysis and interview material further allow us to reason about the underlying mechanics of this behavior.

During the follow-up interviews, participants did express that the preexisting flaws made them less enthusiastic about the task at hand. One went as far as saying that they \say{hardly had the will to finish the task} and merely made sure the tests passed (they were then pleasantly surprised that the system they were assigned in their second task was of significantly higher quality). This is in line with previous findings of~\cite{Besker2020TheMorale} and~\cite{OlssonMeasuringThe}. Interestingly, and seemingly at odds with this statement, the results indicate that debt level does not appear to have significantly affected a subject's propensity to drop out of the experiment.

Considering that the code in a high TD system is both longer and harder to read, we might expect participants to spend more time in those systems. The results indicate no significant effect of TD on the amount of time the participants allocated towards completing their task. Perhaps this is an indication that developers are averse to compensating the productivity loss induced by TD with additional time, resulting in the loss of quality. This is further corroborated by some comments from our interviews, such as \say{I'll just run my tests and be done with it} (when discussing their effort put into the high TD system).

While we found that mimicry alone is not enough to explain the additional TD introduced (Finding 3), it is likely the case that it is a significant driver of TD propagation in general. Given the comparatively large effects on code reuse and variable naming, it is also likely that mimicry is a large part of the effects observed in this study. One participant admitted that \say{you think that someone else thought it should be this way, and then you follow in the same tracks} and several others raised the issue of \emph{code uniformity}. While it may be clear that a system contains issues, if the defects are systematic, it may not necessarily be the case that breaking that uniformity is preferable. For example, using a different (but better) variable name to describe something already existing in adjacent classes could be more confusing than a bad (but consistent) naming scheme. 
We did anticipate this, and it is why we discerned \emph{new} variables from \emph{copied} variables. However, although there is no practical advantage to mimicking the bad naming scheme for \emph{new} variables, it may be that some subjects preferred the \emph{aesthetics} of consistently short and non-descriptive variable names. 

Another takeaway from our result is that developers were at least partially aware of the TD they introduced (Finding 4). \cite{Besker2019SoftwareWork} previously found that developers often felt forced to introduce new TD due to existing TD, which implies an awareness of their TD creation. However, we presented participants with tasks that could feasibly be solved \emph{without} introducing further TD, i.e., they were \emph{not} forced to do so. Our results suggest that developers are somewhat aware of their introduction of TD even under such circumstances. From a practical perspective, this may be good news; it is likely easier to achieve change in a conscious rather than unconscious behavior.

Interestingly, the predictive values of the background factors gathered by our survey were small and uncertain enough that we could safely exclude them from the final models. This, of course, could be a symptom of our relatively low sample size ($N=51$). Years of professional programming experience, which we included as a predictor in the final version of all models measuring TD outcomes, did show a small effect in the expected direction in each case (i.e., more experienced developers introduce less TD on average). However, it is critical to note that we only ever investigated background factors in relation to the amount of TD introduced, but \emph{not} the effect of background factors on susceptibility to BWT effects. It could be that experienced developers are less likely to be affected by existing TD, but that would require the inclusion of \emph{interaction effects} and our sample size would not allow for such a reckless approach.

Finally, although we argue that our results have answered the question of if there is \emph{ever} a BWT effect, the next question may be: \say{is there ever \emph{not} a BWT effect in software engineering?} We suspect that such exceptions are possible, e.g., if a developer is not aware of the existence of a more optimal solution that would reduce their maintenance workload. In other words, could they identify a \emph{broken window}, having never seen one that is \emph{unbroken}? The broken window of the BWT is a message indicating general indifference; it stands to reason that for a \emph{true} BWT effect to occur, the disorder must be apparent to the observer.

Given that we find convincing evidence of BWT effects, companies and software practitioners should continue striving toward keeping the TD density of their systems low. This is by no means a new insight; it has been advocated for by pioneers of the software industry for many years. However, this study is the first to empirically \emph{validate} the claims of the advocates of the BWT and \emph{corroborate} their anecdotal evidence. We believe our results are a significant contribution toward establishing the generalizability of the BWT across disciplines.

\subsection{Suggestions for Future Research}
To isolate and measure the causal effect described by the BWT this study has been designed as a carefully controlled experiment in an artificial setting. While necessary to achieve our goals, a natural continuation of the work and conclusions laid out in this paper would be to do field experiments as well as observational and longitudinal studies along similar lines of inquiry to better understand how this effect is influenced by other factors of software development. Such research would help us define best practices and models to handle technical debt in a software engineering context by better understanding the mechanics influencing its introduction and growth.

At the core of the BWT is that the \emph{broken window} itself is not the primary cause of the undesired BWT effects; it merely communicates an atmosphere of neglect and indifference that indicates that there are no repercussions to the behavior that \say{broke the window}. 

The \emph{state} of the (proverbial) window is just one such indication, although perhaps the most important one. The possibility of repercussions can also be communicated by the \emph{context} of the window. The \emph{broken windows} may still be there, but certain practices and circumstances may inoculate developers against their effects. Examples of such factors could be peer reviewing, continuous integration, and acknowledgment of TD issues (such as a \say{fix me} comment). Examining such contextual factors would be the next major step towards understanding the BWT in software engineering. This could be challenging to accomplish in a controlled experimental setting; it might require a longitudinal field experiment.

Furthermore,~\cite{WILSONGEORGEL.KELLING1982BrokenSafety}, as well as~\cite{Hunt1999TheMaster}, describe a long-term cultural deterioration as a result of BWT effects. While our results support the central mechanism of the BWT, we can not confirm whether or not \say{broken windows} have a \emph{lasting} impact on the culture of a project group, e.g., if a team is assigned to work with high TD systems for some time, will that affect their performance when switching back to a system of low TD density? This is a question we would like to see answered by further research.

Several of the studies that came up during our related literature review (Sect.~\ref{sec:related_literature}) presented models of TD that did not even entertain the possibility of BWT effects. Rather, they exclusively discuss the dynamic of TD \emph{forcing} new TD implementation. We deem our results convincing enough that TD researchers should consider them in future models.

While the BWT was the focus of this study it is only a small part of what may influence a developer in their craft. Further exploring other lines of inquiry relating to developer behavior theory and behavioral design, in the context of software development, will be essential to form effective practices for developer experience.

Finally, we would welcome attempts to replicate the results of our experiment and variations of it using different systems and other examples of TD, perhaps with a sample size sufficient to investigate \emph{interaction effects} and more than two levels of TD density. To facilitate any such efforts, we have made all our research materials publicly available.\footnote{\githubgroup{}}

\subsection{Threats to Validity}
This section discusses threats to the validity of our study and is divided into four parts as per the recommendations of~\cite{Wohlin2012ExperimentationEngineering}

\subsubsection{Internal Validity}
The choice of a simulated experiment allowed us to achieve a high internal quality as we measure the effect of a specific treatment in an environment we control. We also took multiple additional measures to ensure that nothing other than the TD itself influenced the developers to take different actions in the two different tasks they were assigned.

We designed the experiment to ensure that it was always possible to measure the difference between a subject's performance in a high debt system and a low debt system, reducing the influence of factors that we could not control. These factors include, but are not limited to, the time of day of the experiment, which hardware they used, and their physical environment. 

Another threat to internal validity is learning effects. However, randomization of the order of the scenarios and debt levels should have neutralized the impacts of any such effects on our results.

In an isolated experiment such as this, it could have been that the subjects were not particularly meticulous in their work since they did not \emph{care} about their performance. Conversely, in practice, the opposite is more often the case. In experiment and survey research, respondents frequently elicit \emph{social desirability bias}. That is, they may answer or perform in the manner that they think is \emph{socially desirable}~\citep{Phillips1972SomeStudies}. We would argue that since low TD code would be desirable, the subjects are more likely to have \emph{overperformed} rather than the opposite.

The use of a fully automated research tool ensured that we could not influence the subject in any way while participating in the experiment. The research tool presented participants with the same interface, information, and task description no matter whether they were coding in the high debt system or the low debt system. We also denied any questions from participants during their participation.

\subsubsection{External Validity}
Since we had to rely on volunteering subjects, i.e., something more akin to \emph{convenience sampling} rather than random sampling, there is reason to doubt that the sample was fully representative of the population (software developers). However, the fact that background factors showed little predictive power indicates that the skew in the sample is unlikely to have affected our results. 

Another concern regarding external validity is how representative our system and research environments were of real systems and natural environments; there are some disparities, e.g., the lack of symbol searching in the environment and the size of the scenarios. The contrived setting of this study could impact the generalizability of our findings. But, as noted by~\cite{Stol2018TheResearch}, there is an inherent trade-off between the generalizability of a study conducted in a more natural environment and the precision with which measurement of behavior can be achieved in a study using a more controlled and \emph{contrived} setting. The chosen research strategy allowed us to measure how high TD levels affect the behavior of developers, with, we would argue, high precision. This would not have been possible in a natural setting. The qualitative part of the study also helped improve the generalizability of the study as it included a question about BWT in a neutral setting, where multiple participants expressed their support of the theory.

\subsubsection{Construct Validity}
There is a multitude of slightly differing definitions and classifications of TD; hence there may be some questions as to whether our dependent variables really measure TD\@. Similarly, the issues that we introduced to our scenarios might not, by some standard, constitute TD\@. However, we would argue that they do qualify under practically every definition that we have come across while researching the subject. This is further supported by the fact that subjects noticed the defects we had introduced and agreed with our assertion that they lower maintainability, i.e., they rated the \emph{high TD} systems as worse in terms of \say{quality (maintainability)} (Sect.~\ref{sec:result_qual_rating}) and their own work as worse when it had more TD according to the chosen metrics (Sect.~\ref{sec:result_self_rating}). In the follow-up interviews, they more frequently raised the bad variable names as an issue, but the odd participant also noted the code duplication. Taken together, we would argue that this validates our choice of TD representations.

Although we tried to keep our research questions `secret', some participants may have guessed that we were investigating something related to the introduction of TD\@. This could have caused them to act in accordance with their preconceptions of how TD is introduced. While this is not exactly hypothesis guessing, it could have had a similar effect. One of the measures taken to avoid this was the placement of all survey questions relating to TD after both tasks to ensure that those questions could not affect the measurements.

The fact that subjects were able to drop out without completing the tasks could have masked the effects of the debt level. However, the results of our analysis of drop-out behaviors show no significant correlation between the level of debt and a participant's propensity to forgo their assignment. 

\subsubsection{Conclusion Validity}
Given that a relationship was found between a preexisting high TD level and several of the outcomes measuring different kinds of TD using $95\%$ credible intervals, there is reason to believe the validity of our conclusion is high. 

We performed three out of the four types of triangulation suggested by~\cite{Miller2008TriangulationEngineering} to improve the conclusion validity:

\begin{itemize}
    \item \textbf{Methodological triangulation}:
    The cross-validation of results obtained through the application of two distinct research methodologies, one quantitative and one qualitative, improves the validity of our conclusions.
    \item \textbf{Researcher triangulation}:
    Two of the authors performed the manual data extraction processes on all submissions independently. Results were compared, double checked, and any inconsistencies were resolved. This practice helped us reduce any errors or biases introduced by us as researchers.
    \item \textbf{Data triangulation}: The usage of multiple sampling strategies, as well as the collection of data over a span of about $30$ days, ensured some degree of data triangulation. The sampling strategies included snowball sampling where participants were encouraged to pass on the participation invite, convenience sampling by recruiting participants from our personal networks, and purposive sampling by actively seeking out participants with more experience.
\end{itemize}

Several other studies have found, through interviews, that preexisting TD causes new TD (Sect.~\ref{sec:related_literature}). While those findings do not point towards the existence of BWT effects specifically, they could be partially explained by BWT effects. Furthermore, we have found no literature contradicting the BWT in software engineering. In summary, while the literature might not corroborate our findings, it is at least aligned with them.

Any statistical analysis rests on a number of assumptions regarding the underlying distributions. An advantage of \emph{Bayesian data analysis} is that the model design makes those assumptions \emph{explicit}. However, since there is no widely used standard governing this process, subjective assessments could have impacted the results~\citep{Gelman2020BayesianWorkflow}. While we have taken great care to create models that fairly portray the data, we welcome any scrutiny of our process through the examination of our replication package.\footnote{Available at \rppresent{}, source available at \rpgithub{} and \rpdoi{}.}

\section{Conclusions}

For over twenty years, members of the software engineering community have claimed that the \emph{broken windows theory} (BWT) is as applicable to code as its architects would argue it is to crime. This study aimed to empirically evaluate these claims that were previously supported exclusively by anecdotal evidence.

Interpreting \emph{technical debt} (TD) as the metaphorical \emph{broken windows}, we designed an experiment that tasked subjects with extending small \textsf{Java} systems with novel functionality. Unbeknownst to them, we had purposely riddled half of the systems with TD in the form of \emph{bad variable names} and \emph{code duplication}. A group of $29$ participants, some still students, others with more than $30$ years of industry experience, submitted their solutions using a uniform development environment of our design.

From the resulting data set of $51$ distinct solutions, we extracted several measures of TD, some through carefully designed manual procedures and others using the static code analyzing tool \textsf{SonarQube}. Our analysis of these metrics revealed considerable effects of existing TD on the subjects' propensity to reimplement, rather than reuse functionality, choose poor names for their variables, and introduce other issues identified by \textsf{SonarQube}.

The fact that \emph{three} distinct effects were identified, is substantial evidence that the dynamics of the BWT apply to the software development context. The resulting increase in issues of types other than those that we had deliberately introduced suggests that the effects are not exclusively the result of mimicry. Additionally, our finding that subjects tended to rate their work lower when they introduced more TD indicates that they were, at least partially, aware of the shortcuts they had taken.

We further examined the phenomenon through follow-up interviews with volunteering subjects which, through thematic analysis, produced results consistent with the findings of our quantitative analysis.

While our findings lend credence to the claims by programming luminaries like Hunt and Thomas, we can not confirm their assertion that the \emph{long term effects} of the BWT dynamic is, unless actively managed, an inevitable spiral towards oblivion. Nor can we speak to the effects on project culture or the potentially mitigative effects of certain development practices, but we hope that future studies can explore these exciting lines of inquiry.

\section*{Acknowledgments}
We want to thank the participants for their time, Jesper~Olsson for inspiring us to undertake this study, and finally Felicia~Wallin who assisted with the creation of graphical assets.

The computations were enabled by resources provided by the Swedish National Infrastructure for Computing (SNIC), partially funded by the Swedish Research Council through grant agreement no.\ 2018--05973.

Part of this research was funded by Marianne \& Marcus Wallenberg(2017.0071).

\section*{Conflict of Interest}
The authors declared that they have no conflict of interest.

\section*{Data Availability Statement}
This study resulted in multiple digital assets:
\begin{itemize}
    \item The experiment data set.\footnote{Available at \datagithub{} and \datadoi{}}
    \item The experiment scenarios.\footnote{Available at \scenariogithub{} and \scenariodoi{}}
    \item The research tool.\footnote{Available at \rrgithub{} and \rrdoi{}}
    \item The analysis replication package.\footnote{Presented at \rppresent{} with sources available at \rpgithub{} and \rpdoi{}}
\end{itemize}

\backmatter

\bibliography{references}

\end{document}